\documentclass[final,reprint,twocolumn,floatfix,notitlepage,prb]{revtex4-1}
\usepackage{graphicx,amssymb,soul,color,amsmath,bm}
\usepackage{epstopdf,hyperref}
\usepackage{braket,dsfont}
\usepackage[mathcal]{euscript}

\hypersetup{colorlinks=true,citecolor=blue,linkcolor=magenta}

\sethlcolor{yellow}
\allowdisplaybreaks

\begin{document}

\title{The spectral function of Mott-insulating Hubbard ladders: From fractionalized excitations to coherent quasi-particles}
\author{Chun Yang}
\affiliation{Department of Physics, Northeastern University, Boston, Massachusetts 02115, USA}
\author{Adrian E. Feiguin}
\affiliation{Department of Physics, Northeastern University, Boston, Massachusetts 02115, USA}

\date{\today}
\begin{abstract}
We study the spectral function of two-leg Hubbard ladders with the time-dependent density matrix renormalization group method (tDMRG). The high-resolution spectrum displays features of spin-charge separation and a scattering continuum of excitations with coherent bands of bound states ``leaking'' from it. As the inter-leg hopping is increased, the continuum in the bonding channel moves to higher energies and spinon and holon  branches merge into a single coherent quasi-particle band. Simultaneously, the spectrum undergoes a crossover from a regime with two minima at incommensurate values of $k_x$ (a Mott insulator), to one with a single minimum at $k_x=\pi$ (a band insulator). We identify the presence of a continuum of scattering states consisting of a triplon and a polaron. 
We analyze the processes leading to quasiparticle formation by studying the time evolution of charge and spin degrees of freedom in real space after the hole is created. At short times, incoherent holons and spinons are emitted but after a characteristic time $\tau$ charge and spin form polarons that propagate coherently.  
\end{abstract}
\pacs{ 71.30.+h, 71.10.Fd, 74.72.Gh, 79.60.-i}
\maketitle

\section{INTRODUCTION}

The combination of strong interactions and low dimensionality gives rise to exotic and unexpected behavior in quantum many-body systems. In the particular case of fermions in one spatial dimension (1D), the pervasive nesting in the Fermi surfaces (which now consists of just two points at the Fermi level) makes perturbation theory unviable and, as a consequence, Fermi liquid theory breaks down: the natural excitations of the system are described in terms of bosonic modes --holons and spinons--, one carrying the charge quantum number and the other carrying the spin, each with well defined momenta $q_h$ and $q_s$, respectively. An electron with momentum $k$ ``splits'' into holons and spinons, but momentum conservation requires that $k=q_h+q_s$. As a result, the spectrum is characterized by an incoherent continuum of excitations\cite{EsslerBook}. Hence, unlike conventional metals or semiconductors, Fermi quasi-particles are absent. This phenomenon is referred-to as ``spin-charge separation'' and the corresponding low-energy theory as ``Luttinger liquid theory''\cite{GiamarchiBook,Gogolin,Haldane1981}. 
Even though spin-charge separation is intrinsically a manifestation of
1D physics, the possibility of its presence in two-dimensions
(2D) or quasi-2D systems has been extensively debated, particularly
within the context of high-temperature superconductivity\cite{Anderson2000}.
Part of the controversy circles around the interpretation of the pseudogap phase in the cuprates, upon crossing the boundary from the superconducting to the normal state. Instead of a closing of the superconducting gap, experiments \cite{Loeser1997,Shen1995} show a suppression of the quasiparticle peak at the Fermi level. This behavior is difficult to understand in terms of a phase transition and are better interpreted as electrons fractionalizing into charge and spin degrees of freedom in the normal state\cite{Lee1999,Carlson2000}. 
 In addition, it has been suggested that kink or waterfalls observed in photoemission experiments \cite{Damascelli2003} could be attributed to spin-charge separation and traced back adiabatically one-dimensional aspects of the spectrum\cite{Kohno2012,Kohno2015,Yang2016}.
Whether spin-charge separation, or electron-phonon interactions are responsible for the unexpected spectral features in cuprates still is open to interpretation and a topic of great debate.

In this context, much research has been devoted to the study of the Fermi Hubbard Hamiltonian, which has become a paradigmatic model in condensed matter, not only for its relative simplicity, but mainly because it contains the basic ingredients to understand the physics emerging from strong interactions. Moreover, its two-dimensional version has been assumed for decades to be the minimal model to explain high temperature superconductivity\cite{ScalapinoReview,Lee2006} and has acquired even more relevance recently in view of current efforts to realize it in cold atomic systems\cite{Jordens2008,Esslinger2010,Endres2011,Georgescu2014,Hart2015,Duarte2015,Boll2016,Cheuk2016,Brown2017,Hilker2017,Mazurenko2017,Gross2017,Scherg2018,Koepsell2018,Salomon2019,Nichols2019}.
In this work, we consider anisotropic hopings along the legs $t_x$ and along the rungs $t_y$, taking $t_x=1$ as our unit of energy:
\begin{eqnarray}
H = & - & t_x\sum_{i,\lambda,\sigma}\left(c^\dagger_{i,\lambda\sigma} c^{\phantom{\dagger}}_{i+1,\lambda\sigma}+\mathrm{h.c.}\right) + \nonumber\\
& - & t_y\sum_{i,\sigma}\left(c^\dagger_{i2\sigma} c^{\phantom{\dagger}}_{i1\sigma}+\mathrm{h.c.}\right) + U \sum_{i,\lambda} n_{i,\lambda\uparrow}n_{i,\lambda\downarrow},
\end{eqnarray}
where the operator  $c^\dagger_{i\lambda\sigma}$ creates an electron on rung $i$ and leg $\lambda=1,2$ with spin $\sigma=\uparrow,\downarrow$, $n_{i\lambda\sigma}$ is the electron number operator, and $U$ parametrizes the on-site Coulomb repulsion.

Recent results obtained by combining the adaptive time-dependent density matrix renormalization group (tDMRG) method \cite{White2004a,Daley2004,Feiguin2005,Vietri} as a solver for cluster perturbation theory (CPT) \cite{Gross1993,Senechal2000,Senechal2002} indicate that several features associated to spinons and holons survive in the spectral function of the 2D Hubbard model\cite{Yang2016}. These calculations use very large two-leg ladders and are in remarkable agreement with quantum Monte Carlo (QMC)\cite{Bulut1994,Preuss1995,Preuss1997,Grober2000}, variational cluster approximation (VCA) \cite{Dahnken2004,Aichhorn2006}, and dynamical cluster approximation (DCA) \cite{Macridin2006} on square clusters, indicating that Hubbard ladders contain a great deal of information and about the 2D physics. The spectrum shows signatures of both, coherent polaron-like quasiparticles and fractionalization in terms of spinons and holons. Since two-dimensional antiferromagnetic long range order exists only at zero temperature, it is conceivable that the CPT spectrum is a faithful representation of the excitations of the system at finite temperature, after the correlation length reduces to a few lattice spacings, as also suggested by the aforementioned QMC results\cite{Grober2000}.

In a sense, two-leg ladders are a bit pathological: in the Mott insulating phase, both spin and charge degrees of freedom are gapped \cite{Fabrizio1993a,Fabrizio1993b,Castellani1994,Balents1996,Noack1994,Noack1996,Lin1998,Assaraf2004} and spins tend to form rung singlets that condense into a ``rung-singlet phase''. Doping with holes is quite different than doping a two-dimensional antiferromagnet. Upon the introduction of a vacancy (by removing an electron and breaking a singlet), the hole will tend to bind with the unpaired fermion and form a polaron that behaves as a Landau quasi-particle. Two types of polaron exist, corresponding to the symmetric and antisymmetric channels ($k_y=0,\pi$). 

In the Ising-limit corresponding to the $t-J_z$ model\cite{Dagotto1994b,Chernyshev1999,Chernyshev2000,Chernyshev2002,Chernyshev2003,Smakov2007,Smakov2007b}, it is easy to see that the motion of the hole would leave a string of flipped spins behind leading to a linear confining potential. 
Since in reality excitations move in a spin-liquid background, a theoretical treatment becomes complicated. However, in the strong $t_y$ limit, the hole moves in a vacuum of rung dimers and both, theoretical and numerical approaches offer good agreement corroborating the presence polaronic quasi-particles \cite{Eder1998,Sushkov1999,Endres1996,Brunner2001}. In the weak coupling regime, theory is based on bosonization and RG arguments\cite{Fabrizio1993a,Fabrizio1993b,Balents1996,Lin1998} and also supports the quasi-particle picture. 

The transition between 1D-like physics and coherent polaron-like quasiparticles is not easily identifiable, since the binding energy between a holon and a spinon, or the quasi-particle weight, are hard to measure quantities.  
One could qualitatively anticipate possible scenarios: If the binding energy is too small, the hole may find it energetically favorable to move along the leg direction. The resulting physics will be mainly one-dimensional and the excitations will consist of deconfined holons and spinons. As the binding energy increases, a coherent band of bound states will ``leak'' from the spinon-holon continuum and they may become the lowest energy excitations. 
Hence, this could be interpreted as a two particle problem, in which both could propagate independently, or as composite bound state. 

Some numerical studies in this
direction, looking at 2,3 and 4-leg $t-J$ ladders, indicate the presence
of spinon and holon excitations \cite{Poilblanc1995,Haas1996a,Rice1997,Martins2000,Brunner2001}.
Recently, a series of works proposed \cite{Zhu2013,Zhu2014,Zhu2015,Zhu2015b,Zhu2016,Zhu2018} that doping a Mott insulating $t-J$ ladder would result in localization of the hole and, as a consequence, the system would not support conventional quasi-particles. In a subsequent study, White {\it et al.} \cite{White2015} demonstrated using extensive numerical calculations that in reality there is no localization but a change in the quasiparticle dispersion, with the minimum of the hole band moving away from $k_x=\pi$ as the ratio $\alpha=t_x/t_y$ is varied\cite{Troyer1996,Riera1999} . Therefore, doping the Mott insulating ladder would be equivalent to doping a band-insulator and dressed holes would form robust quasi-particles. Large scale DMRG studies \cite{Liu2016} confirm this picture, in which the quasi-particle mass diverges at a critical value of the anisotropy parameter $\alpha$.  
Authors argue that in the large $\alpha$ regime, where the chains are weakly coupled, the polaron is an extended object with a complex internal structure in which charge and spin locally behave as separate degrees of freedom. 

In order to shed light on these questions, we carry out time-dependent DMRG simulations \cite{White2004a,Daley2004,Feiguin2005,Vietri} that allow us to obtain spectra with unprecedented resolution. We present an analysis of the results for the excitation spectrum in section \ref{results}, together with a study of the charge and spin dynamics in real time to identify the nature of the processes leading to quasi particle formation. We conclude with a summary and discussion of our findings. 

\begin{figure*}
 \center
  \includegraphics[width=\textwidth]{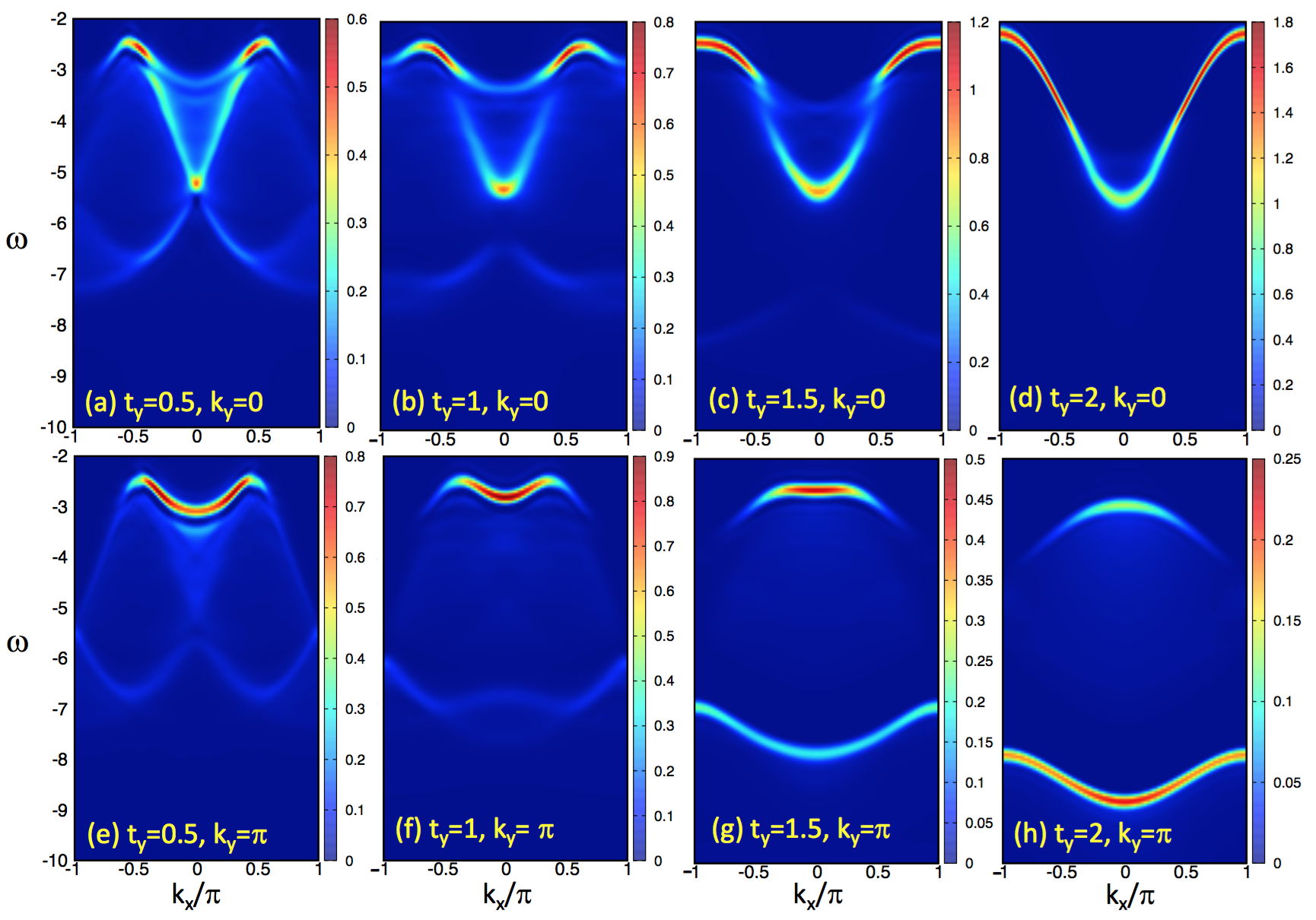}
\caption{Spectral function of a Hubbard ladder with $L=80$ and $U/t=8$, at half-filling, obtained with tDMRG. Top and bottom rows show the symmetric and anti-symmetric sectors, respectively.
We only plot the photo-emission part of the spectrum. Notice that the color scale varies from panel to panel.
}
  \label{fig:arpes}
\end{figure*}

\section{RESULTS}\label{results}
\subsection{Spectral function}

We have calculated the photoemission spectrum of a single hole for $2 \times L$ Mott insulating Hubbard ladders with $L=80$ using the adaptive time-dependent DMRG method (Notice that the inverse photoemission spectrum is simply related  by a particle-hole transformation). 
 We used a time step $dt=0.02$ and up to 800 DMRG states, that for times $t < 40$ translates into a truncation error of the order of $10^-5$ or smaller (larger errors correspond to small values of $t_y$).
 This technique has been extensively described in the literature and we refer the reader to Refs.~[\onlinecite{Feiguin2005,Vietri}] for details. The single particle Green's function $G^<(x,t)=\langle c^\dagger_x(t) c_x(0) \rangle$ is measured in real-time and space, and Fourier transformed to frequency using a Hann window with $t_{max}=40$ in order to minimize boundary effects and other artifacts such as ringing resulting from the finite size of the lattice and time interval. We have not found it necessary to use the linear prediction introduced in Ref.[\onlinecite{Barthel2009}].

\begin{figure}
  \includegraphics[width=0.45\textwidth]{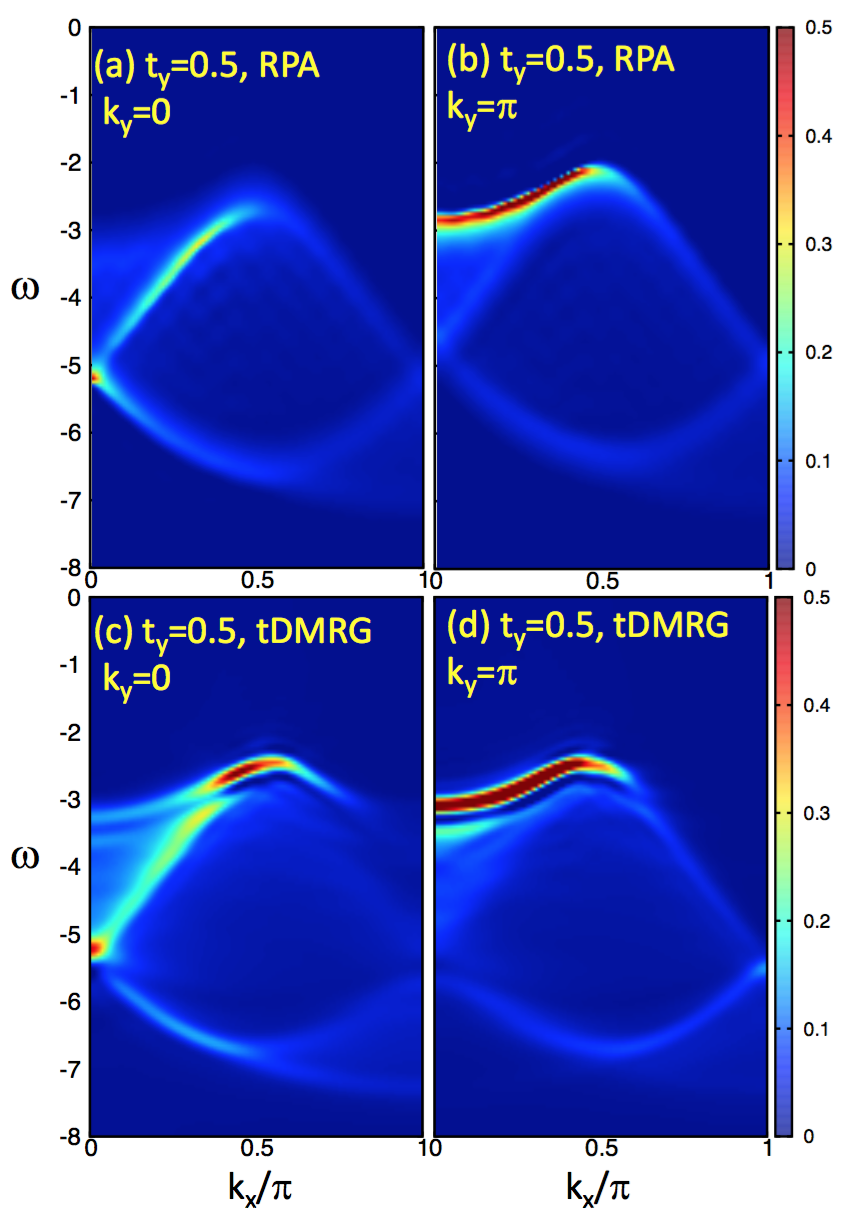}
\caption{RPA spectral function for the ladder with $U=8$, $t_y=0.5$ obtained from the single chain results for (a) symmetric ($k_y=0$) and (b) anti-symmetric ($k_y=\pi$) sectors. (c) and (d) show the corresponding spectra from tDMRG simulations on ladders. }
  \label{fig:rpa}
\end{figure}

 Results for the spectral function are shown in Fig.\ref{fig:arpes}, where the color density depicts the spectral weight as a function of momentum $k_x$ and frequency $\omega$. Each column corresponds to different values of $t_y$ and  each row to two possible transverse momenta $k_y=0,\pi$ representing even and odd, or bonding and anti-bonding symmetry sectors with respect to reflections along the leg direction.
For small $t_y$ we find clear signatures of spin charge separation. Curiously, most of the spectral weight on the spinon branch goes to the $k_y=\pi$ sector, while the holon branches dominate the $k_y=0$ spectrum. In Figs.\ref{fig:arpes}(a) and (e) we clearly see an avoided level crossing at low energies that indicates mixing between spin and charge, with the spectral weight accumulating around the Fermi points. 
At energies larger than the effective $J_y\sim t_y^2/U$, the polaron would not be well defined and the hole would move without an associated spin degree of freedom, same as a holon in 1D chains.    
As $t_y$ increases, we find another avoided level crossing at $k_x=0$ that merge the two holon branches into a single band with finite curvature. This resembles the spectrum of a single hole in a spin-incoherent Luttinger liquid\cite{Fiete2004,Cheianov2004,Cheianov2005,Fiete2006,Fiete2007b,Halperin2007,Feiguin2009d,Nocera2018}, in which the vacancy moves in a background of incoherent spins.  

These results can be compared to those obtained from an RPA treatment of the single chain spectrum. It is calculated by solving the equation $G^{-1}=G_0^{-1}-t_y \cos{(k_y)}$, where $G_0$ is the exact Green's function of the 1D Hubbard chain and $k_y=0,\pi$. The resulting spectral function $A(k_x,\omega)=-\frac{1}{\pi}G(k_x,\omega)$ for $t_y=0.5$ is shown in Fig.\ref{fig:rpa} and presents some of the same features as Figs.~\ref{fig:arpes}(a)-(f) reproduced in Figs.~\ref{fig:rpa}((c) and (d), namely, the dominant holon and spinon branches in the $k_y=0$ and $\pi$ sectors, respectively. This behavior is simply explained by the structure of the RPA solution: the imaginary part of the Green's function contains a contribution from the real part of $G0$, which has branch cuts and can change sign. On the other hand, we notice the absence of bound states, which are however expected from the field theoretical analysis of the problem\cite{Essler2002}. This is probably due to the relatively small value of $t_y$ and the resolution of our numerical spectrum that may hinder their observation.

The large $t_y$ regime (Fig.\ref{fig:arpes}(d) and (h)) is intuitively easier to understand: the Mott insulating ground state is a product of local rung dimers . Upon doping with a single hole, a coherent plane wave of rung polarons is created on top of the dimer vacuum. This is equivalent to introducing a vacancy in a chain of spinless fermions with one particle per site (a band insulator), and leads to a cosine-like dispersion. However, after paying careful attention to the processes taking place during the polaron motion, one realizes that a local bound state of a particle and a hole cannot move without introducing fluctuations in the spin background \cite{Eder1998,Sushkov1999}. This effect is stronger for small $t_y$, giving rise to an effective second neighbor hopping and, depending on the parameters of the problem, the minimum of the hole dispersion can shift away from $k_x=\pi$, as observed in Fig.\ref{fig:arpes}(b). 

In order to seek stronger support for this physical picture, we construct a polaron variational wave function following Ref.~[\onlinecite{Endres1996}]. In the so-called ``Local Rung Approximation'' (LRA), the Mott insulating state $|\psi_0\rangle$ consists of a product of localized single rung dimers $|S_i\rangle$, each being the the ground state of the local rung Hamiltonian (the exact ground states for $t_x=0$). 
The excited states are constructed as plane waves with a single polaron, that can assume two possible values $k_y=0,\pi$:
\begin{eqnarray}
|\psi_1(k_x,k_y)\rangle = \sum_{x} e^{ik_x x} |x,k_y\rangle,
\end{eqnarray}
where the state $|x\rangle$ is defined as
\begin{equation}
|x,k_y\rangle = |S_1\rangle|S_2\rangle\cdots|k_y\rangle\cdots |S_L\rangle.
\end{equation}
The dispersion for $\omega < 0$ is given by:
\begin{eqnarray}
\omega(k_x,k_y=0,\pi) & = & \langle \psi_0|H|\psi_0\rangle \nonumber \\
&-& \langle \psi_1(k_x,k_y)|H|\psi_1(k_x,k_y)\rangle-U/2 \\
& = & E_0+U/2+ t_y e^{ik_y} -tA(k_y)\cos{(k)}, \nonumber
\end{eqnarray}
with $A(k_y)=(1+e^{ik_y}E_1/2t_y)^2/(1+E_1^2/4t_y^2)^2$ and $E_{0,1}=-U/2\mp \sqrt{(U/2)^2+4t_y^2}$.

\begin{figure}
 \center
  \includegraphics[width=0.45\textwidth]{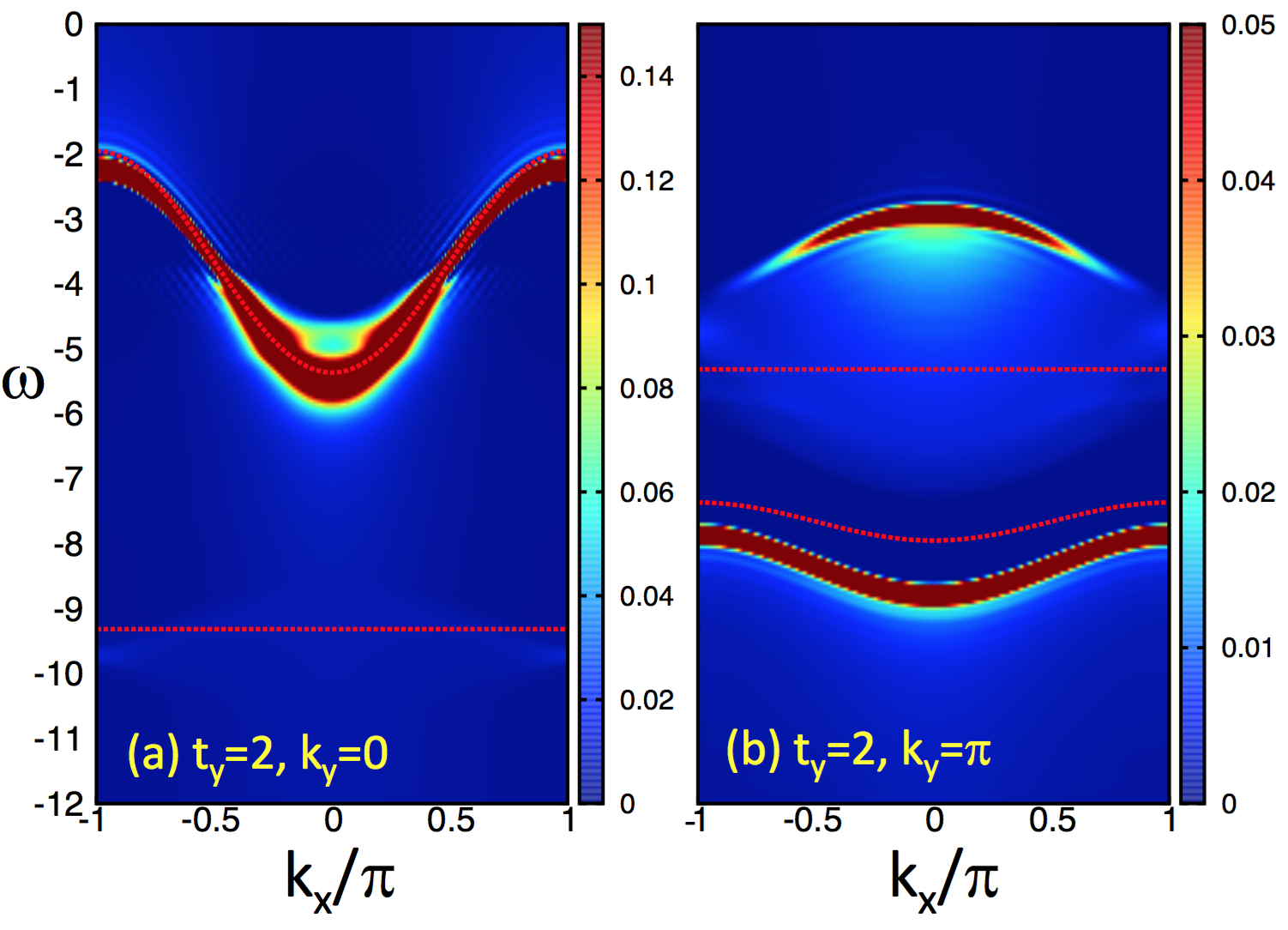}
\caption{Results for $t_y=2$ compared to the local rung approximation (LRA) in dashed red lines (see text). This color scale makes the two particle continua visible. }
  \label{fig:lra}
\end{figure}

\begin{figure*}
 \center
  \includegraphics[width=\textwidth]{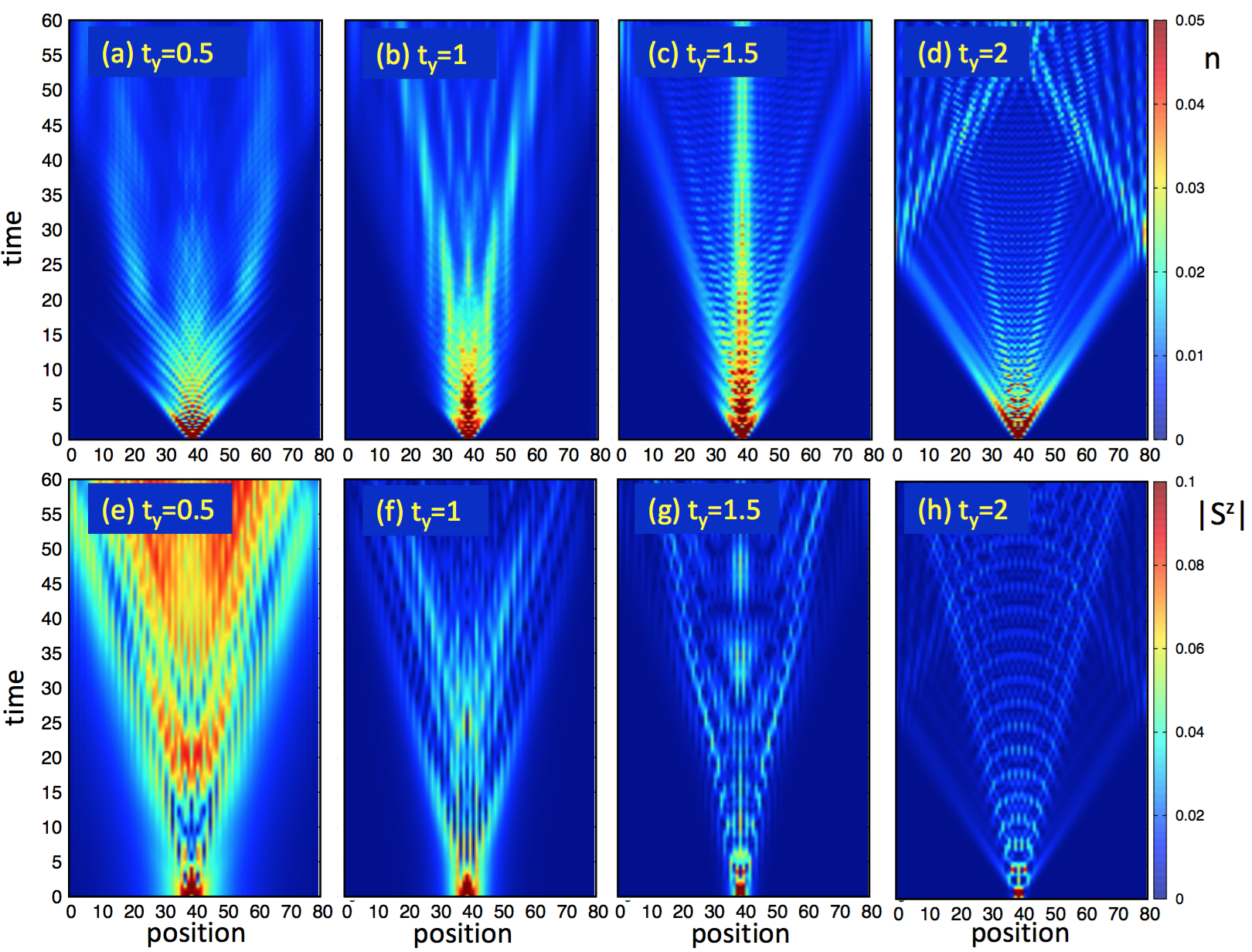}
\caption{Top panels display the time evolution of charge density after a hole is introduced at the center of one of the legs. Similar results for the spin density are plotted in the bottom pannels. We only show data for the leg on which the hole is created. }
  \label{fig:lightcone}
\end{figure*}

In Fig.~\ref{fig:lra} we reproduce the results for $t_y=2$ in a different color scale to resolve fainter features in the spectrum. We first observe that the coherent band in the bonding sector is perfectly described by the LRA, as previously reported in QMC calculations\cite{Endres1996,Brunner2001}. However, In the $k_y=\pi$ channel, the LRA yields energies slightly lower, while correctly describing the corrections to the bandwidth. The smaller bandwidth results from cancellations due to the symmetry of the wave-function and the fermionic sign that introduce destructive interference preventing precesses that do not conserve double occupation. This also translates into a much smaller spectral weight in this band, which makes it difficult to resolve with other numerical methods\cite{Endres1996}.
Our tDMRG calculations allow us to identify a continuum near the bottom of the bonding band and a weaker one at energies centered around $\sim -10t$. In the anti-bonding sector, we observe another continuum of excitations at {\it low} energies of the order of $-5t$. The high intensity peak at low energies in Fig.\ref{fig:lra}(b) corresponds to the edge of a two-particle continuum. In the symmetric sector, we attribute the high energy states to a triplon and an anti-bonding polaron with $k_y=\pi$, while in the anti-symmetric sector, to a triplon and a bonding polaron with $k_y=0$.
The zeroth order energies of these states within the LRA are $\omega_+= 2E_0-U/2-t_y$ and $\omega_-=2E_0-U/2+t_y$. Therefore, the scattering states of a triplon and a symmetric polaron live in the anti-symmetric sector and vice-versa. In the weak coupling case, the scattering continuum for $k_y=0$ overlaps with the polaron band. As the interchain hopping increases, the continuum moves to higher energies and the coherent dispersive band becomes wider in momentum. Similar effect occurs in the $k_y=\pi$ sector, but with the scattering continuum shifting to lower energies. 
The sharp edge of the continuum in the anti-bonding sector could be interpreted as unstable bound states of a triplon and a polaron with higher spin $S=3/2$ or a ``spin bag''\cite{Schrieffer1988,Schrieffer1989}, as suggested in Ref.~\onlinecite{Sushkov1999}]. This is confirmed by exact diagonalization results on ladders of sizes up to $2 \times 7$ (not shown). The $S=3/2$ states are more robust near $k_x=0$ while states with $S=1/2$ appear at higher energies and toward the edge of the Brillouin zone. 
Notice that a triplon can assume three possible polarizations $|\uparrow\uparrow\rangle,|\downarrow,\downarrow\rangle$, and $(|\uparrow\downarrow\rangle+|\downarrow\uparrow\rangle)/\sqrt{2}$, and the symmetric polaron $(|\sigma,0\rangle-|0,\sigma\rangle)/\sqrt{2}$ can have two. This means that one can pair a $Sz=1$ triplon and $\sigma=\downarrow$ polaron, or $Sz=0$ and $\sigma=\uparrow$ polaron.
Even though this continuum is not necessarily a signature of spin charge separation, the fact that the polaron may not have a well defined spin polarization could be interpreted as a behavior that is typically associated to a holon, carrying charge but not spin. 
As $t_y$ is increased, the coherent band in the anti-bonding channel is pushed to higher energies and its spectral weight is reduced. In this regime the system effectively becomes a single band insulator. Remarkably, the bandwidth of the bonding band is $\sim 4t$ meaning that polarons can propagate coherently through {\it first} order processes. This does not occur for the anti-bonding polaron, which explains the limitations of LRA in this sector. 

 It is important to highlight some outstanding differences with the commonly discussed $t-J$ model. In that case, double occupancy is forbidden and the band insulating regime is interpreted as polarons moving in a background of spin singlets\cite{Noack1996,Troyer1996}. In order for polarons to be able to hop coherently, second order processes (hopping plus spin-flip) are required, which is reflected in a considerably reduced bandwidth \cite{Eder1998,Sushkov1999}. In the Hubbard model, this reduced bandwidth is observed for values of the interaction $U \gg t_y$ (not shown).

\subsection{Real-time dynamics}

In order to confirm this picture we carried out a ``time-of-flight'' numerical experiment by creating a vacancy at the center of the ladder and observing the propagation of the density $\langle n_i(t)\rangle$ and spin $\langle S^z_i(t) \rangle$ fluctuations, as displayed in Fig.\ref{fig:lightcone}. For simplicity we show only results for one of the legs where the vacancy is created. We notice nodes along the $x$ direction that result from the density alternating between legs. In 1D chains (not shown here) one observes \cite{Jagla1993,Kollath2005,Cheneau2012,Herbrych2017} two lightcones of excitations propagating coherently with maximum velocity $v_s$ and $v_c$ for spin and charge, respectively. We focus our attention on panels Fig.\ref{fig:lightcone}(a) and (e) corresponding to small $t_y=0.5$. At short times we also see two lightcones that propagate with the characteristic spin and charge velocities. However, the emitted holons fade away rapidly, with the wave packet spreading over the entire volume and losing coherence, while the spinons remain coherent up to the largest simulated time. At longer times $t\sim 15$ , we see the emergence of two clearly defined branches in panel (a) that have the same slope as the spinons. The picture is now clearer: after injecting a vacancy, incoherent holons and spinons are emitted, but after a characteristic time $\tau \sim 1/t_y$ a polaron is formed, that propagates with a velocity $v_p \sim v_s$. As $t_y$ is increased, holons become heavier and, for large $t_y$, polarons are the only type of excitation that remain observable. In this case we see two dominant branches corresponding to the maximum velocities for the $k_y=0$ and $k_y=\pi$ coherent bands. Interestingly, for $t_y=1.5$ we find a clear and bright mode that seems localized: this is simply due to the curvature of the dispersion for $k_y=0$, which becomes practically flat. It was shown that the effective mass diverges at the value of $1/\alpha=t_y/t_x \sim 1.4$ for $U/t=8$ \cite{White2015,Liu2016}. This type of localization should not be associated to a breakdown of the Fermi-liquid picture, as argued in Refs.~[\onlinecite{Zhu2013,Zhu2014,Zhu2015,Zhu2015b,Zhu2016,Zhu2018}].

\begin{figure}
 \center
  \includegraphics[width=0.45\textwidth]{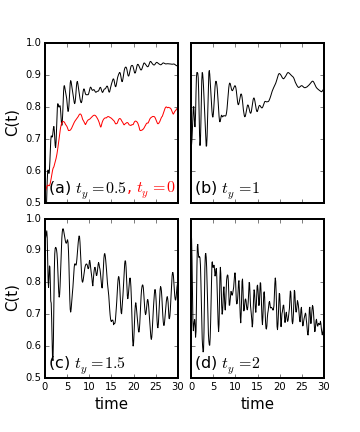}
\caption{Cross correlation between charge and spin density as a function of time, for different values of $t_y$. In panel (a) we include results for a single chain with $U=8$ for comparison.}
  \label{fig:cross}
\end{figure}

In order to establish a measure of coherence between both spin and charge degrees of freedom, we analyze the cross correlation between the two datasets. For each time slice we calculate the quantity:
\begin{equation}
C(t)=\frac{\sum_i \langle n_{i1}(t) \rangle |\langle S^z_{i2}(t) \rangle|}{\sqrt{\sum_i \langle n_{i1}(t) \rangle ^2}\sqrt{\sum_i \langle S^z_{i2}(t) \rangle ^2} }.
\end{equation}
Notice that this is equivalent to the overlap between two normalized vectors, one with components defined by the density on one leg $\langle n_{i1}(t)\rangle$ and the other, by the spin on the second leg of the ladder, $|\langle S^z_i \rangle|$. If the two quantities are perfectly correlated, $C=1$. 
Results for the cross correlation are displayed in Fig.\ref{fig:cross} for different values of $t_y$. In the Mott insulating regime for $t_y=0.5, 1$ we observe rapid oscillations preceding a quasi-steady behavior. We associate the transient to the time scale required for the polaron to form, $\tau$. We show results for $t_y=0$ (single chain) for comparison. The fact that the correlation reaches a finite value is not necessarily a sign of correlation between charge ans spin: consider for instance an idealized scenario in which the charge and spin densities are uniformly distributed within their respective lightcones; it is easy to realize that the cross correlation function would saturate to a value $C=\sqrt{(v_c v_s)/v_s}$. In the band insulating regime for $t_y=1.5, 2$ the behavior is actually more complex due to the presence of two clearly distinct lightcones corresponding to the $k_y=0$ and $\pi$ channels that produce a great deal of interference, which is enhanced by the rapid oscillations of the hole along the rung in the transverse direction. In this case, spin and charge are quite correlated but most of the overlap is concentrated in the $k_y=0$ sector that contributes with the greater weight. Ideally one would like to resolve and compare the contributions of both lightcones separately, that should therefore be normalized independently. Since there is no obvious way to do this, we find that the overall correlation is actually reduced. For $t_y=2$ we are no longer able to clearly distinguish a transient, although we can identify a dip that is associated to the wave packets bouncing off the edges of the ladder. Interestingly, although we cannot assert this with clarity, for weak interchain coupling we see indications that $\tau$ and, consequently, the polaron binding energy do not depend strongly on $t_y$ and (at least in the Mott insulating regime and for this value of $U$) is of the order of $\tau \sim 20$. This time seems considerably reduced after the system undergoes a crossover to the band insulating regime.


\section{Conclusions}

Our tDMRG simulations offer a new perspective on the physics o Hubbard ladders allowing us to resolve fine details of the spectrum with unprecedented resolution. Among some of the main features we highlight the appearance of avoided level crossings at weak coupling indicating hybridization between spin and charge, although most of the main features of the spectrum can still be traced back to the physics of one-dimensional chains. Moreover, we find that the excitation spectrum is dominated by multi-particle scattering states. Coherent polarons emerge from this continua as spin $S=1/2$ and charge $e$ quasi particles, a bound state of a holon and a spinon. Symmetric $k_y=0$ polarons can propagate coherently through first order processes and become the relevant excitations in the large $t_y$ regime. On the other hand, anti-symmetric polarons become heavier and lose spectral weight due to the symmetry of the wave-function. We find that the most important scattering states consist of a triplon and a polaron, which can assume different spin polarizations. Interestingly, scattering between triplons and symmetric(antisymmetric) polarons are responsible for the continuum in the $k_y=\pi(0)$ sector, and the coherent quasi-particles emerge from these continua, as the scattering states shift to lower(higher) energies.  
In addition, our results clearly demonstrate the importance of charge fluctuations and, in particular, accounting for processes involving double occupancy for properly describing the coherent propagation of quasi-particles when $U$ is not too large.

In the anti-symmetric channel, the continuum of scattering states has a sharp edge at low energies that consists of states with spin $S=3/2$.
These states with higher spin near the $(0,\pi)$ point implies that ''spin bags`` might be realized in the two-dimensional counterpart between the $X$ and $M$ points of the Brillouin zone. These excitations would have a short lifetime and decay into a triplon with $S^z=0$ or $1$ and a polaron with spin $S^z=1/2$ or $-1/2$, respectively. This is a manifestation of the Nagaoka mechanism\cite{Nagaoka1966} in which a hole surrounds itself by ferromagnetic cloud to propagate more efficiently\cite{Schrieffer1988,Schrieffer1989,Troyer1996,Kohno1997}.

In order for the spin polaron to behave as coherent quasi-particle, the system size has to be considerably larger than the characteristic size of the polaron $\chi$. 
If $\chi$ is of the order or larger than the system size, numerical results would not be able to resolve the quasi-particle and would mistakenly lead one to the conclusion that quasi-particles are not stable objects. The proper interpretation would be that the system flows, in the RG sense, toward the strong rung coupling limit in which the polaron, at long wave-lengths (or, as seen from afar), is a well-defined quasi-particle \cite{Fabrizio1993a,Fabrizio1993b,Balents1996,Lin1998}. This seems so confirm old speculation about the nature of the single hole doped ground state in ladders\cite{Martins1999,Martins2000a,Martins2000b}.

In some aspects, these arguments are similar to those in the single-impurity Kondo problem: in the strong coupling limit a magnetic impurity becomes a scattering center and the problem can be elegantly described in terms of Fermi-liquid theory\cite{Nozieres1985}. 
However, in finite systems and at intermediate couplings, only the internal structure of the Kondo state can be resolved \cite{Yang2016}. 
Instead of a magnetic impurity, in our case we deal with a mobile impurity (a spinon) that is dressed by a holon (or vice-versa). 

At small interchain hopping, it is reasonable to assume that a small finite temperature will overcome the binding that holds the polaron-like quasiparticle together and only deconfined holons and spinons would survive\cite{Grober2000,Raczkowski2013}. Therefore, unless the binding energy is large enough, or the temperature small enough, experiments are unlikely to be able to resolve sharp quasi-particle features. 

\acknowledgements
The authors thank the NSF for support through Grant No. DMR-1807814.
We are grateful to M. Fabrizio, G. Sierra, F. Essler, A. Stepanyants, R. Pereira and A. Chernyshev for illuminating discussions.


\begin{thebibliography}{101}%
\makeatletter
\providecommand \@ifxundefined [1]{%
 \@ifx{#1\undefined}
}%
\providecommand \@ifnum [1]{%
 \ifnum #1\expandafter \@firstoftwo
 \else \expandafter \@secondoftwo
 \fi
}%
\providecommand \@ifx [1]{%
 \ifx #1\expandafter \@firstoftwo
 \else \expandafter \@secondoftwo
 \fi
}%
\providecommand \natexlab [1]{#1}%
\providecommand \enquote  [1]{``#1''}%
\providecommand \bibnamefont  [1]{#1}%
\providecommand \bibfnamefont [1]{#1}%
\providecommand \citenamefont [1]{#1}%
\providecommand \href@noop [0]{\@secondoftwo}%
\providecommand \href [0]{\begingroup \@sanitize@url \@href}%
\providecommand \@href[1]{\@@startlink{#1}\@@href}%
\providecommand \@@href[1]{\endgroup#1\@@endlink}%
\providecommand \@sanitize@url [0]{\catcode `\\12\catcode `\$12\catcode
  `\&12\catcode `\#12\catcode `\^12\catcode `\_12\catcode `\%12\relax}%
\providecommand \@@startlink[1]{}%
\providecommand \@@endlink[0]{}%
\providecommand \url  [0]{\begingroup\@sanitize@url \@url }%
\providecommand \@url [1]{\endgroup\@href {#1}{\urlprefix }}%
\providecommand \urlprefix  [0]{URL }%
\providecommand \Eprint [0]{\href }%
\providecommand \doibase [0]{http://dx.doi.org/}%
\providecommand \selectlanguage [0]{\@gobble}%
\providecommand \bibinfo  [0]{\@secondoftwo}%
\providecommand \bibfield  [0]{\@secondoftwo}%
\providecommand \translation [1]{[#1]}%
\providecommand \BibitemOpen [0]{}%
\providecommand \bibitemStop [0]{}%
\providecommand \bibitemNoStop [0]{.\EOS\space}%
\providecommand \EOS [0]{\spacefactor3000\relax}%
\providecommand \BibitemShut  [1]{\csname bibitem#1\endcsname}%
\let\auto@bib@innerbib\@empty
\bibitem [{\citenamefont {Essler}\ \emph {et~al.}(2010)\citenamefont {Essler},
  \citenamefont {Frahm}, \citenamefont {G\"ohmann}, \citenamefont {Kl\"umper},\
  and\ \citenamefont {Korepin}}]{EsslerBook}%
  \BibitemOpen
  \bibfield  {author} {\bibinfo {author} {\bibfnamefont {F.}~\bibnamefont
  {Essler}}, \bibinfo {author} {\bibfnamefont {H.}~\bibnamefont {Frahm}},
  \bibinfo {author} {\bibfnamefont {F.}~\bibnamefont {G\"ohmann}}, \bibinfo
  {author} {\bibfnamefont {A.}~\bibnamefont {Kl\"umper}}, \ and\ \bibinfo
  {author} {\bibfnamefont {V.~E.}\ \bibnamefont {Korepin}},\ }\href@noop {}
  {\emph {\bibinfo {title} {The One-Dimensional Hubbard Model}}}\ (\bibinfo
  {publisher} {Cambridge University Press, Cambridge, England},\ \bibinfo
  {year} {2010})\BibitemShut {NoStop}%
\bibitem [{\citenamefont {Giamarchi}(2004)}]{GiamarchiBook}%
  \BibitemOpen
  \bibfield  {author} {\bibinfo {author} {\bibfnamefont {T.}~\bibnamefont
  {Giamarchi}},\ }\href@noop {} {\emph {\bibinfo {title} {Quantum Physics in
  One Dimension}}}\ (\bibinfo  {publisher} {Clarendon Press, Oxford},\ \bibinfo
  {year} {2004})\BibitemShut {NoStop}%
\bibitem [{\citenamefont {Gogolin}\ \emph {et~al.}(1998)\citenamefont
  {Gogolin}, \citenamefont {Nerseyan},\ and\ \citenamefont
  {Tsvelik}}]{Gogolin}%
  \BibitemOpen
  \bibfield  {author} {\bibinfo {author} {\bibfnamefont {A.~O.}\ \bibnamefont
  {Gogolin}}, \bibinfo {author} {\bibfnamefont {A.~A.}\ \bibnamefont
  {Nerseyan}}, \ and\ \bibinfo {author} {\bibfnamefont {A.~M.}\ \bibnamefont
  {Tsvelik}},\ }\href@noop {} {\emph {\bibinfo {title} {Bosonization and
  Strongly Correlated Systems}}}\ (\bibinfo  {publisher} {Cambridge University
  Press, Cambridge, England},\ \bibinfo {year} {1998})\BibitemShut {NoStop}%
\bibitem [{\citenamefont {Haldane}(1981)}]{Haldane1981}%
  \BibitemOpen
  \bibfield  {author} {\bibinfo {author} {\bibfnamefont {F.~D.~M.}\
  \bibnamefont {Haldane}},\ }\href@noop {} {\bibfield  {journal} {\bibinfo
  {journal} {J. Phys. C}\ }\textbf {\bibinfo {volume} {14}},\ \bibinfo {pages}
  {2585} (\bibinfo {year} {1981})}\BibitemShut {NoStop}%
\bibitem [{\citenamefont {Anderson}(2000)}]{Anderson2000}%
  \BibitemOpen
  \bibfield  {author} {\bibinfo {author} {\bibfnamefont {P.~W.}\ \bibnamefont
  {Anderson}},\ }\href@noop {} {\bibfield  {journal} {\bibinfo  {journal}
  {Phys. C Supercond.}\ }\textbf {\bibinfo {volume} {341-348}},\ \bibinfo
  {pages} {9} (\bibinfo {year} {2000})}\BibitemShut {NoStop}%
\bibitem [{\citenamefont {Loeser}\ \emph {et~al.}(1997)\citenamefont {Loeser},
  \citenamefont {Shen}, \citenamefont {Schabel}, \citenamefont {Kim},
  \citenamefont {Zhang}, \citenamefont {Kapitulnik},\ and\ \citenamefont
  {Fournier}}]{Loeser1997}%
  \BibitemOpen
  \bibfield  {author} {\bibinfo {author} {\bibfnamefont {A.~G.}\ \bibnamefont
  {Loeser}}, \bibinfo {author} {\bibfnamefont {Z.-X.}\ \bibnamefont {Shen}},
  \bibinfo {author} {\bibfnamefont {M.~C.}\ \bibnamefont {Schabel}}, \bibinfo
  {author} {\bibfnamefont {C.}~\bibnamefont {Kim}}, \bibinfo {author}
  {\bibfnamefont {M.}~\bibnamefont {Zhang}}, \bibinfo {author} {\bibfnamefont
  {A.}~\bibnamefont {Kapitulnik}}, \ and\ \bibinfo {author} {\bibfnamefont
  {P.}~\bibnamefont {Fournier}},\ }\href {\doibase 10.1103/PhysRevB.56.14185}
  {\bibfield  {journal} {\bibinfo  {journal} {Phys. Rev. B}\ }\textbf {\bibinfo
  {volume} {56}},\ \bibinfo {pages} {14185} (\bibinfo {year}
  {1997})}\BibitemShut {NoStop}%
\bibitem [{\citenamefont {Shen}\ and\ \citenamefont {Dessau}(1995)}]{Shen1995}%
  \BibitemOpen
  \bibfield  {author} {\bibinfo {author} {\bibfnamefont {Z.-X.}\ \bibnamefont
  {Shen}}\ and\ \bibinfo {author} {\bibfnamefont {D.}~\bibnamefont {Dessau}},\
  }\href {\doibase https://doi.org/10.1016/0370-1573(95)80001-A} {\bibfield
  {journal} {\bibinfo  {journal} {Physics Reports}\ }\textbf {\bibinfo {volume}
  {253}},\ \bibinfo {pages} {1 } (\bibinfo {year} {1995})}\BibitemShut
  {NoStop}%
\bibitem [{\citenamefont {Lee}(1999)}]{Lee1999}%
  \BibitemOpen
  \bibfield  {author} {\bibinfo {author} {\bibfnamefont {P.~A.}\ \bibnamefont
  {Lee}},\ }\href {\doibase https://doi.org/10.1016/S0921-4534(99)00059-3}
  {\bibfield  {journal} {\bibinfo  {journal} {Physica C: Superconductivity}\
  }\textbf {\bibinfo {volume} {317}},\ \bibinfo {pages} {194 } (\bibinfo {year}
  {1999})}\BibitemShut {NoStop}%
\bibitem [{\citenamefont {Carlson}\ \emph {et~al.}(2000)\citenamefont
  {Carlson}, \citenamefont {Orgad}, \citenamefont {Kivelson},\ and\
  \citenamefont {Emery}}]{Carlson2000}%
  \BibitemOpen
  \bibfield  {author} {\bibinfo {author} {\bibfnamefont {E.~W.}\ \bibnamefont
  {Carlson}}, \bibinfo {author} {\bibfnamefont {D.}~\bibnamefont {Orgad}},
  \bibinfo {author} {\bibfnamefont {S.~A.}\ \bibnamefont {Kivelson}}, \ and\
  \bibinfo {author} {\bibfnamefont {V.~J.}\ \bibnamefont {Emery}},\ }\href
  {\doibase 10.1103/PhysRevB.62.3422} {\bibfield  {journal} {\bibinfo
  {journal} {Phys. Rev. B}\ }\textbf {\bibinfo {volume} {62}},\ \bibinfo
  {pages} {3422} (\bibinfo {year} {2000})}\BibitemShut {NoStop}%
\bibitem [{\citenamefont {Damascelli}\ \emph {et~al.}(2003)\citenamefont
  {Damascelli}, \citenamefont {Hussain},\ and\ \citenamefont
  {Shen}}]{Damascelli2003}%
  \BibitemOpen
  \bibfield  {author} {\bibinfo {author} {\bibfnamefont {A.}~\bibnamefont
  {Damascelli}}, \bibinfo {author} {\bibfnamefont {Z.}~\bibnamefont {Hussain}},
  \ and\ \bibinfo {author} {\bibfnamefont {Z.-X.}\ \bibnamefont {Shen}},\
  }\href@noop {} {\bibfield  {journal} {\bibinfo  {journal} {Rev. Mod. Phys.}\
  }\textbf {\bibinfo {volume} {75}},\ \bibinfo {pages} {473} (\bibinfo {year}
  {2003})}\BibitemShut {NoStop}%
\bibitem [{\citenamefont {Kohno}(2012)}]{Kohno2012}%
  \BibitemOpen
  \bibfield  {author} {\bibinfo {author} {\bibfnamefont {M.}~\bibnamefont
  {Kohno}},\ }\href@noop {} {\bibfield  {journal} {\bibinfo  {journal} {Phys.
  Rev. Lett.}\ }\textbf {\bibinfo {volume} {108}},\ \bibinfo {pages} {076401}
  (\bibinfo {year} {2012})}\BibitemShut {NoStop}%
\bibitem [{\citenamefont {Kohno}(2015)}]{Kohno2015}%
  \BibitemOpen
  \bibfield  {author} {\bibinfo {author} {\bibfnamefont {M.}~\bibnamefont
  {Kohno}},\ }\href@noop {} {\bibfield  {journal} {\bibinfo  {journal} {Phys.
  Rev. B}\ }\textbf {\bibinfo {volume} {92}},\ \bibinfo {pages} {085129}
  (\bibinfo {year} {2015})}\BibitemShut {NoStop}%
\bibitem [{\citenamefont {Yang}\ and\ \citenamefont
  {Feiguin}(2016)}]{Yang2016}%
  \BibitemOpen
  \bibfield  {author} {\bibinfo {author} {\bibfnamefont {C.}~\bibnamefont
  {Yang}}\ and\ \bibinfo {author} {\bibfnamefont {A.~E.}\ \bibnamefont
  {Feiguin}},\ }\href@noop {} {\bibfield  {journal} {\bibinfo  {journal} {Phys.
  Rev. B}\ }\textbf {\bibinfo {volume} {93}},\ \bibinfo {pages} {081107}
  (\bibinfo {year} {2016})}\BibitemShut {NoStop}%
\bibitem [{\citenamefont {Scalapino}(2007)}]{ScalapinoReview}%
  \BibitemOpen
  \bibfield  {author} {\bibinfo {author} {\bibfnamefont {D.}~\bibnamefont
  {Scalapino}},\ }in\ \href {\doibase 10.1007/978-0-387-68734-6_13} {\emph
  {\bibinfo {booktitle} {Handbook of High-Temperature Superconductivity}}},\
  \bibinfo {editor} {edited by\ \bibinfo {editor} {\bibfnamefont
  {J.}~\bibnamefont {Schrieffer}}\ and\ \bibinfo {editor} {\bibfnamefont
  {J.}~\bibnamefont {Brooks}}}\ (\bibinfo  {publisher} {Springer New York},\
  \bibinfo {year} {2007})\ pp.\ \bibinfo {pages} {495--526}\BibitemShut
  {NoStop}%
\bibitem [{\citenamefont {Lee}\ \emph {et~al.}(2006)\citenamefont {Lee},
  \citenamefont {Nagaosa},\ and\ \citenamefont {Wen}}]{Lee2006}%
  \BibitemOpen
  \bibfield  {author} {\bibinfo {author} {\bibfnamefont {P.~A.}\ \bibnamefont
  {Lee}}, \bibinfo {author} {\bibfnamefont {N.}~\bibnamefont {Nagaosa}}, \ and\
  \bibinfo {author} {\bibfnamefont {X.-G.}\ \bibnamefont {Wen}},\ }\href
  {\doibase 10.1103/RevModPhys.78.17} {\bibfield  {journal} {\bibinfo
  {journal} {Rev. Mod. Phys.}\ }\textbf {\bibinfo {volume} {78}},\ \bibinfo
  {pages} {17} (\bibinfo {year} {2006})}\BibitemShut {NoStop}%
\bibitem [{\citenamefont {Jordens}\ \emph {et~al.}(2008)\citenamefont
  {Jordens}, \citenamefont {Strohmaier}, \citenamefont {Gunter}, \citenamefont
  {Moritz},\ and\ \citenamefont {Esslinger}}]{Jordens2008}%
  \BibitemOpen
  \bibfield  {author} {\bibinfo {author} {\bibfnamefont {R.}~\bibnamefont
  {Jordens}}, \bibinfo {author} {\bibfnamefont {N.}~\bibnamefont {Strohmaier}},
  \bibinfo {author} {\bibfnamefont {K.}~\bibnamefont {Gunter}}, \bibinfo
  {author} {\bibfnamefont {H.}~\bibnamefont {Moritz}}, \ and\ \bibinfo {author}
  {\bibfnamefont {T.}~\bibnamefont {Esslinger}},\ }\href@noop {} {\bibfield
  {journal} {\bibinfo  {journal} {Nature}\ }\textbf {\bibinfo {volume} {455}},\
  \bibinfo {pages} {204} (\bibinfo {year} {2008})}\BibitemShut {NoStop}%
\bibitem [{\citenamefont {Esslinger}(2010)}]{Esslinger2010}%
  \BibitemOpen
  \bibfield  {author} {\bibinfo {author} {\bibfnamefont {T.}~\bibnamefont
  {Esslinger}},\ }\href {\doibase 10.1146/annurev-conmatphys-070909-104059}
  {\bibfield  {journal} {\bibinfo  {journal} {Annual Review of Condensed Matter
  Physics}\ }\textbf {\bibinfo {volume} {1}},\ \bibinfo {pages} {129} (\bibinfo
  {year} {2010})},\ \Eprint
  {http://arxiv.org/abs/https://doi.org/10.1146/annurev-conmatphys-070909-104059}
  {https://doi.org/10.1146/annurev-conmatphys-070909-104059} \BibitemShut
  {NoStop}%
\bibitem [{\citenamefont {Endres}\ \emph {et~al.}(2011)\citenamefont {Endres},
  \citenamefont {Cheneau}, \citenamefont {Fukuhara}, \citenamefont
  {Weitenberg}, \citenamefont {Schau{\ss}}, \citenamefont {Gross},
  \citenamefont {Mazza}, \citenamefont {Ba{\~n}uls}, \citenamefont {Pollet},
  \citenamefont {Bloch},\ and\ \citenamefont {Kuhr}}]{Endres2011}%
  \BibitemOpen
  \bibfield  {author} {\bibinfo {author} {\bibfnamefont {M.}~\bibnamefont
  {Endres}}, \bibinfo {author} {\bibfnamefont {M.}~\bibnamefont {Cheneau}},
  \bibinfo {author} {\bibfnamefont {T.}~\bibnamefont {Fukuhara}}, \bibinfo
  {author} {\bibfnamefont {C.}~\bibnamefont {Weitenberg}}, \bibinfo {author}
  {\bibfnamefont {P.}~\bibnamefont {Schau{\ss}}}, \bibinfo {author}
  {\bibfnamefont {C.}~\bibnamefont {Gross}}, \bibinfo {author} {\bibfnamefont
  {L.}~\bibnamefont {Mazza}}, \bibinfo {author} {\bibfnamefont {M.~C.}\
  \bibnamefont {Ba{\~n}uls}}, \bibinfo {author} {\bibfnamefont
  {L.}~\bibnamefont {Pollet}}, \bibinfo {author} {\bibfnamefont
  {I.}~\bibnamefont {Bloch}}, \ and\ \bibinfo {author} {\bibfnamefont
  {S.}~\bibnamefont {Kuhr}},\ }\href {\doibase 10.1126/science.1209284}
  {\bibfield  {journal} {\bibinfo  {journal} {Science}\ }\textbf {\bibinfo
  {volume} {334}},\ \bibinfo {pages} {200} (\bibinfo {year} {2011})},\ \Eprint
  {http://arxiv.org/abs/http://science.sciencemag.org/content/334/6053/200.full.pdf}
  {http://science.sciencemag.org/content/334/6053/200.full.pdf} \BibitemShut
  {NoStop}%
\bibitem [{\citenamefont {Georgescu}\ \emph {et~al.}(2014)\citenamefont
  {Georgescu}, \citenamefont {Ashhab},\ and\ \citenamefont
  {Nori}}]{Georgescu2014}%
  \BibitemOpen
  \bibfield  {author} {\bibinfo {author} {\bibfnamefont {I.}~\bibnamefont
  {Georgescu}}, \bibinfo {author} {\bibfnamefont {S.}~\bibnamefont {Ashhab}}, \
  and\ \bibinfo {author} {\bibfnamefont {F.}~\bibnamefont {Nori}},\ }\href@noop
  {} {\bibfield  {journal} {\bibinfo  {journal} {Rev. Mod. Phys.}\ }\textbf
  {\bibinfo {volume} {86}},\ \bibinfo {pages} {153} (\bibinfo {year}
  {2014})}\BibitemShut {NoStop}%
\bibitem [{\citenamefont {Hart}\ \emph {et~al.}(2015)\citenamefont {Hart},
  \citenamefont {Duarte}, \citenamefont {Yang}, \citenamefont {Liu},
  \citenamefont {Paiva}, \citenamefont {Khatami}, \citenamefont {Scalettar},
  \citenamefont {Trivedi}, \citenamefont {Huse},\ and\ \citenamefont
  {Hulet}}]{Hart2015}%
  \BibitemOpen
  \bibfield  {author} {\bibinfo {author} {\bibfnamefont {R.~A.}\ \bibnamefont
  {Hart}}, \bibinfo {author} {\bibfnamefont {P.~M.}\ \bibnamefont {Duarte}},
  \bibinfo {author} {\bibfnamefont {T.-L.}\ \bibnamefont {Yang}}, \bibinfo
  {author} {\bibfnamefont {X.}~\bibnamefont {Liu}}, \bibinfo {author}
  {\bibfnamefont {T.}~\bibnamefont {Paiva}}, \bibinfo {author} {\bibfnamefont
  {E.}~\bibnamefont {Khatami}}, \bibinfo {author} {\bibfnamefont {R.~T.}\
  \bibnamefont {Scalettar}}, \bibinfo {author} {\bibfnamefont {N.}~\bibnamefont
  {Trivedi}}, \bibinfo {author} {\bibfnamefont {D.~A.}\ \bibnamefont {Huse}}, \
  and\ \bibinfo {author} {\bibfnamefont {R.~G.}\ \bibnamefont {Hulet}},\ }\href
  {https://doi.org/10.1038/nature14223} {\bibfield  {journal} {\bibinfo
  {journal} {Nature}\ }\textbf {\bibinfo {volume} {519}},\ \bibinfo {pages}
  {211 EP } (\bibinfo {year} {2015})}\BibitemShut {NoStop}%
\bibitem [{\citenamefont {Duarte}\ \emph {et~al.}(2015)\citenamefont {Duarte},
  \citenamefont {Hart}, \citenamefont {Yang}, \citenamefont {Liu},
  \citenamefont {Paiva}, \citenamefont {Khatami}, \citenamefont {Scalettar},
  \citenamefont {Trivedi},\ and\ \citenamefont {Hulet}}]{Duarte2015}%
  \BibitemOpen
  \bibfield  {author} {\bibinfo {author} {\bibfnamefont {P.~M.}\ \bibnamefont
  {Duarte}}, \bibinfo {author} {\bibfnamefont {R.~A.}\ \bibnamefont {Hart}},
  \bibinfo {author} {\bibfnamefont {T.-L.}\ \bibnamefont {Yang}}, \bibinfo
  {author} {\bibfnamefont {X.}~\bibnamefont {Liu}}, \bibinfo {author}
  {\bibfnamefont {T.}~\bibnamefont {Paiva}}, \bibinfo {author} {\bibfnamefont
  {E.}~\bibnamefont {Khatami}}, \bibinfo {author} {\bibfnamefont {R.~T.}\
  \bibnamefont {Scalettar}}, \bibinfo {author} {\bibfnamefont {N.}~\bibnamefont
  {Trivedi}}, \ and\ \bibinfo {author} {\bibfnamefont {R.~G.}\ \bibnamefont
  {Hulet}},\ }\href {\doibase 10.1103/PhysRevLett.114.070403} {\bibfield
  {journal} {\bibinfo  {journal} {Phys. Rev. Lett.}\ }\textbf {\bibinfo
  {volume} {114}},\ \bibinfo {pages} {070403} (\bibinfo {year}
  {2015})}\BibitemShut {NoStop}%
\bibitem [{\citenamefont {Boll}\ \emph {et~al.}(2016)\citenamefont {Boll},
  \citenamefont {Hilker}, \citenamefont {Salomon}, \citenamefont {Omran},
  \citenamefont {Nespolo}, \citenamefont {Pollet}, \citenamefont {Bloch},\ and\
  \citenamefont {Gross}}]{Boll2016}%
  \BibitemOpen
  \bibfield  {author} {\bibinfo {author} {\bibfnamefont {M.}~\bibnamefont
  {Boll}}, \bibinfo {author} {\bibfnamefont {T.~A.}\ \bibnamefont {Hilker}},
  \bibinfo {author} {\bibfnamefont {G.}~\bibnamefont {Salomon}}, \bibinfo
  {author} {\bibfnamefont {A.}~\bibnamefont {Omran}}, \bibinfo {author}
  {\bibfnamefont {J.}~\bibnamefont {Nespolo}}, \bibinfo {author} {\bibfnamefont
  {L.}~\bibnamefont {Pollet}}, \bibinfo {author} {\bibfnamefont
  {I.}~\bibnamefont {Bloch}}, \ and\ \bibinfo {author} {\bibfnamefont
  {C.}~\bibnamefont {Gross}},\ }\href {\doibase 10.1126/science.aag1635}
  {\bibfield  {journal} {\bibinfo  {journal} {Science}\ }\textbf {\bibinfo
  {volume} {353}},\ \bibinfo {pages} {1257} (\bibinfo {year} {2016})},\ \Eprint
  {http://arxiv.org/abs/http://science.sciencemag.org/content/353/6305/1257.full.pdf}
  {http://science.sciencemag.org/content/353/6305/1257.full.pdf} \BibitemShut
  {NoStop}%
\bibitem [{\citenamefont {Cheuk}\ \emph {et~al.}(2016)\citenamefont {Cheuk},
  \citenamefont {Nichols}, \citenamefont {Lawrence}, \citenamefont {Okan},
  \citenamefont {Zhang}, \citenamefont {Khatami}, \citenamefont {Trivedi},
  \citenamefont {Paiva}, \citenamefont {Rigol},\ and\ \citenamefont
  {Zwierlein}}]{Cheuk2016}%
  \BibitemOpen
  \bibfield  {author} {\bibinfo {author} {\bibfnamefont {L.~W.}\ \bibnamefont
  {Cheuk}}, \bibinfo {author} {\bibfnamefont {M.~A.}\ \bibnamefont {Nichols}},
  \bibinfo {author} {\bibfnamefont {K.~R.}\ \bibnamefont {Lawrence}}, \bibinfo
  {author} {\bibfnamefont {M.}~\bibnamefont {Okan}}, \bibinfo {author}
  {\bibfnamefont {H.}~\bibnamefont {Zhang}}, \bibinfo {author} {\bibfnamefont
  {E.}~\bibnamefont {Khatami}}, \bibinfo {author} {\bibfnamefont
  {N.}~\bibnamefont {Trivedi}}, \bibinfo {author} {\bibfnamefont
  {T.}~\bibnamefont {Paiva}}, \bibinfo {author} {\bibfnamefont
  {M.}~\bibnamefont {Rigol}}, \ and\ \bibinfo {author} {\bibfnamefont {M.~W.}\
  \bibnamefont {Zwierlein}},\ }\href {\doibase 10.1126/science.aag3349}
  {\bibfield  {journal} {\bibinfo  {journal} {Science}\ }\textbf {\bibinfo
  {volume} {353}},\ \bibinfo {pages} {1260} (\bibinfo {year} {2016})},\ \Eprint
  {http://arxiv.org/abs/http://science.sciencemag.org/content/353/6305/1260.full.pdf}
  {http://science.sciencemag.org/content/353/6305/1260.full.pdf} \BibitemShut
  {NoStop}%
\bibitem [{\citenamefont {Brown}\ \emph {et~al.}(2017)\citenamefont {Brown},
  \citenamefont {Mitra}, \citenamefont {Guardado-Sanchez}, \citenamefont
  {Schau{\ss}}, \citenamefont {Kondov}, \citenamefont {Khatami}, \citenamefont
  {Paiva}, \citenamefont {Trivedi}, \citenamefont {Huse},\ and\ \citenamefont
  {Bakr}}]{Brown2017}%
  \BibitemOpen
  \bibfield  {author} {\bibinfo {author} {\bibfnamefont {P.~T.}\ \bibnamefont
  {Brown}}, \bibinfo {author} {\bibfnamefont {D.}~\bibnamefont {Mitra}},
  \bibinfo {author} {\bibfnamefont {E.}~\bibnamefont {Guardado-Sanchez}},
  \bibinfo {author} {\bibfnamefont {P.}~\bibnamefont {Schau{\ss}}}, \bibinfo
  {author} {\bibfnamefont {S.~S.}\ \bibnamefont {Kondov}}, \bibinfo {author}
  {\bibfnamefont {E.}~\bibnamefont {Khatami}}, \bibinfo {author} {\bibfnamefont
  {T.}~\bibnamefont {Paiva}}, \bibinfo {author} {\bibfnamefont
  {N.}~\bibnamefont {Trivedi}}, \bibinfo {author} {\bibfnamefont {D.~A.}\
  \bibnamefont {Huse}}, \ and\ \bibinfo {author} {\bibfnamefont {W.~S.}\
  \bibnamefont {Bakr}},\ }\href {\doibase 10.1126/science.aam7838} {\bibfield
  {journal} {\bibinfo  {journal} {Science}\ }\textbf {\bibinfo {volume}
  {357}},\ \bibinfo {pages} {1385} (\bibinfo {year} {2017})},\ \Eprint
  {http://arxiv.org/abs/http://science.sciencemag.org/content/357/6358/1385.full.pdf}
  {http://science.sciencemag.org/content/357/6358/1385.full.pdf} \BibitemShut
  {NoStop}%
\bibitem [{\citenamefont {Hilker}\ \emph {et~al.}(2017)\citenamefont {Hilker},
  \citenamefont {Salomon}, \citenamefont {Grusdt}, \citenamefont {Omran},
  \citenamefont {Boll}, \citenamefont {Demler}, \citenamefont {Bloch},\ and\
  \citenamefont {Gross}}]{Hilker2017}%
  \BibitemOpen
  \bibfield  {author} {\bibinfo {author} {\bibfnamefont {T.~A.}\ \bibnamefont
  {Hilker}}, \bibinfo {author} {\bibfnamefont {G.}~\bibnamefont {Salomon}},
  \bibinfo {author} {\bibfnamefont {F.}~\bibnamefont {Grusdt}}, \bibinfo
  {author} {\bibfnamefont {A.}~\bibnamefont {Omran}}, \bibinfo {author}
  {\bibfnamefont {M.}~\bibnamefont {Boll}}, \bibinfo {author} {\bibfnamefont
  {E.}~\bibnamefont {Demler}}, \bibinfo {author} {\bibfnamefont
  {I.}~\bibnamefont {Bloch}}, \ and\ \bibinfo {author} {\bibfnamefont
  {C.}~\bibnamefont {Gross}},\ }\href {\doibase 10.1126/science.aam8990}
  {\bibfield  {journal} {\bibinfo  {journal} {Science}\ }\textbf {\bibinfo
  {volume} {357}},\ \bibinfo {pages} {484} (\bibinfo {year} {2017})},\ \Eprint
  {http://arxiv.org/abs/http://science.sciencemag.org/content/357/6350/484.full.pdf}
  {http://science.sciencemag.org/content/357/6350/484.full.pdf} \BibitemShut
  {NoStop}%
\bibitem [{\citenamefont {Mazurenko}\ \emph {et~al.}(2017)\citenamefont
  {Mazurenko}, \citenamefont {Chiu}, \citenamefont {Ji}, \citenamefont
  {Parsons}, \citenamefont {Kanász-Nagy}, \citenamefont {Schmidt},
  \citenamefont {Grusdt}, \citenamefont {Demler}, \citenamefont {Greif},\ and\
  \citenamefont {Greiner}}]{Mazurenko2017}%
  \BibitemOpen
  \bibfield  {author} {\bibinfo {author} {\bibfnamefont {A.}~\bibnamefont
  {Mazurenko}}, \bibinfo {author} {\bibfnamefont {C.~S.}\ \bibnamefont {Chiu}},
  \bibinfo {author} {\bibfnamefont {G.}~\bibnamefont {Ji}}, \bibinfo {author}
  {\bibfnamefont {M.~F.}\ \bibnamefont {Parsons}}, \bibinfo {author}
  {\bibfnamefont {M.}~\bibnamefont {Kanász-Nagy}}, \bibinfo {author}
  {\bibfnamefont {R.}~\bibnamefont {Schmidt}}, \bibinfo {author} {\bibfnamefont
  {F.}~\bibnamefont {Grusdt}}, \bibinfo {author} {\bibfnamefont
  {E.}~\bibnamefont {Demler}}, \bibinfo {author} {\bibfnamefont
  {D.}~\bibnamefont {Greif}}, \ and\ \bibinfo {author} {\bibfnamefont
  {M.}~\bibnamefont {Greiner}},\ }\href@noop {} {\bibfield  {journal} {\bibinfo
   {journal} {Nature}\ }\textbf {\bibinfo {volume} {545}},\ \bibinfo {pages}
  {462} (\bibinfo {year} {2017})}\BibitemShut {NoStop}%
\bibitem [{\citenamefont {Gross}\ and\ \citenamefont
  {Bloch}(2017)}]{Gross2017}%
  \BibitemOpen
  \bibfield  {author} {\bibinfo {author} {\bibfnamefont {C.}~\bibnamefont
  {Gross}}\ and\ \bibinfo {author} {\bibfnamefont {I.}~\bibnamefont {Bloch}},\
  }\href {\doibase 10.1126/science.aal3837} {\bibfield  {journal} {\bibinfo
  {journal} {Science}\ }\textbf {\bibinfo {volume} {357}},\ \bibinfo {pages}
  {995} (\bibinfo {year} {2017})},\ \Eprint
  {http://arxiv.org/abs/http://science.sciencemag.org/content/357/6355/995.full.pdf}
  {http://science.sciencemag.org/content/357/6355/995.full.pdf} \BibitemShut
  {NoStop}%
\bibitem [{\citenamefont {Scherg}\ \emph {et~al.}(2018)\citenamefont {Scherg},
  \citenamefont {Kohlert}, \citenamefont {Herbrych}, \citenamefont {Stolpp},
  \citenamefont {Bordia}, \citenamefont {Schneider}, \citenamefont
  {Heidrich-Meisner}, \citenamefont {Bloch},\ and\ \citenamefont
  {Aidelsburger}}]{Scherg2018}%
  \BibitemOpen
  \bibfield  {author} {\bibinfo {author} {\bibfnamefont {S.}~\bibnamefont
  {Scherg}}, \bibinfo {author} {\bibfnamefont {T.}~\bibnamefont {Kohlert}},
  \bibinfo {author} {\bibfnamefont {J.}~\bibnamefont {Herbrych}}, \bibinfo
  {author} {\bibfnamefont {J.}~\bibnamefont {Stolpp}}, \bibinfo {author}
  {\bibfnamefont {P.}~\bibnamefont {Bordia}}, \bibinfo {author} {\bibfnamefont
  {U.}~\bibnamefont {Schneider}}, \bibinfo {author} {\bibfnamefont
  {F.}~\bibnamefont {Heidrich-Meisner}}, \bibinfo {author} {\bibfnamefont
  {I.}~\bibnamefont {Bloch}}, \ and\ \bibinfo {author} {\bibfnamefont
  {M.}~\bibnamefont {Aidelsburger}},\ }\href {\doibase
  10.1103/PhysRevLett.121.130402} {\bibfield  {journal} {\bibinfo  {journal}
  {Phys. Rev. Lett.}\ }\textbf {\bibinfo {volume} {121}},\ \bibinfo {pages}
  {130402} (\bibinfo {year} {2018})}\BibitemShut {NoStop}%
\bibitem [{\citenamefont {Koepsell}\ \emph {et~al.}()\citenamefont {Koepsell},
  \citenamefont {Vijayan}, \citenamefont {Sompet}, \citenamefont {Grusdt},
  \citenamefont {Hilker}, \citenamefont {Demler}, \citenamefont {Salomon},
  \citenamefont {Bloch},\ and\ \citenamefont {Gross}}]{Koepsell2018}%
  \BibitemOpen
  \bibfield  {author} {\bibinfo {author} {\bibfnamefont {J.}~\bibnamefont
  {Koepsell}}, \bibinfo {author} {\bibfnamefont {J.}~\bibnamefont {Vijayan}},
  \bibinfo {author} {\bibfnamefont {P.}~\bibnamefont {Sompet}}, \bibinfo
  {author} {\bibfnamefont {F.}~\bibnamefont {Grusdt}}, \bibinfo {author}
  {\bibfnamefont {T.~A.}\ \bibnamefont {Hilker}}, \bibinfo {author}
  {\bibfnamefont {E.}~\bibnamefont {Demler}}, \bibinfo {author} {\bibfnamefont
  {G.}~\bibnamefont {Salomon}}, \bibinfo {author} {\bibfnamefont
  {I.}~\bibnamefont {Bloch}}, \ and\ \bibinfo {author} {\bibfnamefont
  {C.}~\bibnamefont {Gross}},\ }\href@noop {} {\ }\Eprint
  {http://arxiv.org/abs/1811.06907} {arXiv:1811.06907
  [arXiv:cond-mat.quant-gas]} \BibitemShut {NoStop}%
\bibitem [{\citenamefont {Salomon}\ \emph {et~al.}(2019)\citenamefont
  {Salomon}, \citenamefont {Koepsell}, \citenamefont {Vijayan}, \citenamefont
  {Hilker}, \citenamefont {Nespolo}, \citenamefont {Pollet}, \citenamefont
  {Bloch},\ and\ \citenamefont {Gross}}]{Salomon2019}%
  \BibitemOpen
  \bibfield  {author} {\bibinfo {author} {\bibfnamefont {G.}~\bibnamefont
  {Salomon}}, \bibinfo {author} {\bibfnamefont {J.}~\bibnamefont {Koepsell}},
  \bibinfo {author} {\bibfnamefont {J.}~\bibnamefont {Vijayan}}, \bibinfo
  {author} {\bibfnamefont {T.~A.}\ \bibnamefont {Hilker}}, \bibinfo {author}
  {\bibfnamefont {J.}~\bibnamefont {Nespolo}}, \bibinfo {author} {\bibfnamefont
  {L.}~\bibnamefont {Pollet}}, \bibinfo {author} {\bibfnamefont
  {I.}~\bibnamefont {Bloch}}, \ and\ \bibinfo {author} {\bibfnamefont
  {C.}~\bibnamefont {Gross}},\ }\href@noop {} {\bibfield  {journal} {\bibinfo
  {journal} {Nature}\ }\textbf {\bibinfo {volume} {565}},\ \bibinfo {pages}
  {56} (\bibinfo {year} {2019})}\BibitemShut {NoStop}%
\bibitem [{\citenamefont {Nichols}\ \emph {et~al.}(2019)\citenamefont
  {Nichols}, \citenamefont {Cheuk}, \citenamefont {Okan}, \citenamefont
  {Hartke}, \citenamefont {Mendez}, \citenamefont {Senthil}, \citenamefont
  {Khatami}, \citenamefont {Zhang},\ and\ \citenamefont
  {Zwierlein}}]{Nichols2019}%
  \BibitemOpen
  \bibfield  {author} {\bibinfo {author} {\bibfnamefont {M.~A.}\ \bibnamefont
  {Nichols}}, \bibinfo {author} {\bibfnamefont {L.~W.}\ \bibnamefont {Cheuk}},
  \bibinfo {author} {\bibfnamefont {M.}~\bibnamefont {Okan}}, \bibinfo {author}
  {\bibfnamefont {T.~R.}\ \bibnamefont {Hartke}}, \bibinfo {author}
  {\bibfnamefont {E.}~\bibnamefont {Mendez}}, \bibinfo {author} {\bibfnamefont
  {T.}~\bibnamefont {Senthil}}, \bibinfo {author} {\bibfnamefont
  {E.}~\bibnamefont {Khatami}}, \bibinfo {author} {\bibfnamefont
  {H.}~\bibnamefont {Zhang}}, \ and\ \bibinfo {author} {\bibfnamefont {M.~W.}\
  \bibnamefont {Zwierlein}},\ }\href {\doibase 10.1126/science.aat4387}
  {\bibfield  {journal} {\bibinfo  {journal} {Science}\ }\textbf {\bibinfo
  {volume} {363}},\ \bibinfo {pages} {383} (\bibinfo {year} {2019})},\ \Eprint
  {http://arxiv.org/abs/http://science.sciencemag.org/content/363/6425/383.full.pdf}
  {http://science.sciencemag.org/content/363/6425/383.full.pdf} \BibitemShut
  {NoStop}%
\bibitem [{\citenamefont {White}\ and\ \citenamefont
  {Feiguin}(2004)}]{White2004a}%
  \BibitemOpen
  \bibfield  {author} {\bibinfo {author} {\bibfnamefont {S.~R.}\ \bibnamefont
  {White}}\ and\ \bibinfo {author} {\bibfnamefont {A.~E.}\ \bibnamefont
  {Feiguin}},\ }\href@noop {} {\bibfield  {journal} {\bibinfo  {journal} {Phys.
  Rev. Lett.}\ }\textbf {\bibinfo {volume} {93}},\ \bibinfo {pages} {076401}
  (\bibinfo {year} {2004})}\BibitemShut {NoStop}%
\bibitem [{\citenamefont {Daley}\ \emph {et~al.}(2004)\citenamefont {Daley},
  \citenamefont {Kollath}, \citenamefont {Schollw\"ock}, ,\ and\ \citenamefont
  {Vidal}}]{Daley2004}%
  \BibitemOpen
  \bibfield  {author} {\bibinfo {author} {\bibfnamefont {A.~J.}\ \bibnamefont
  {Daley}}, \bibinfo {author} {\bibfnamefont {C.}~\bibnamefont {Kollath}},
  \bibinfo {author} {\bibfnamefont {U.}~\bibnamefont {Schollw\"ock}}, , \ and\
  \bibinfo {author} {\bibfnamefont {G.}~\bibnamefont {Vidal}},\ }\href@noop {}
  {\bibfield  {journal} {\bibinfo  {journal} {J. Stat. Mech.: Theor. Exp.}\ ,\
  \bibinfo {pages} {P04005}} (\bibinfo {year} {2004})}\BibitemShut {NoStop}%
\bibitem [{\citenamefont {Feiguin}\ and\ \citenamefont
  {White}(2005)}]{Feiguin2005}%
  \BibitemOpen
  \bibfield  {author} {\bibinfo {author} {\bibfnamefont {A.}~\bibnamefont
  {Feiguin}}\ and\ \bibinfo {author} {\bibfnamefont {S.}~\bibnamefont
  {White}},\ }\href@noop {} {\bibfield  {journal} {\bibinfo  {journal} {Phys.
  Rev. B}\ }\textbf {\bibinfo {volume} {72}},\ \bibinfo {pages} {20404}
  (\bibinfo {year} {2005})}\BibitemShut {NoStop}%
\bibitem [{\citenamefont {Feiguin}(2011)}]{Vietri}%
  \BibitemOpen
  \bibfield  {author} {\bibinfo {author} {\bibfnamefont {A.~E.}\ \bibnamefont
  {Feiguin}},\ }in\ \href@noop {} {\emph {\bibinfo {booktitle} {XV Training
  Course in the Physics of Strongly Correlated Systems}}},\ Vol.\ \bibinfo
  {volume} {1419}\ (\bibinfo  {publisher} {AIP Proceedings},\ \bibinfo {year}
  {2011})\ p.~\bibinfo {pages} {5}\BibitemShut {NoStop}%
\bibitem [{\citenamefont {Gros}\ and\ \citenamefont
  {Valenti}(1993)}]{Gross1993}%
  \BibitemOpen
  \bibfield  {author} {\bibinfo {author} {\bibfnamefont {C.}~\bibnamefont
  {Gros}}\ and\ \bibinfo {author} {\bibfnamefont {R.}~\bibnamefont {Valenti}},\
  }\href@noop {} {\bibfield  {journal} {\bibinfo  {journal} {Phys. Rev. B}\
  }\textbf {\bibinfo {volume} {48}},\ \bibinfo {pages} {418} (\bibinfo {year}
  {1993})}\BibitemShut {NoStop}%
\bibitem [{\citenamefont {S\'en\'echal}\ \emph {et~al.}(2000)\citenamefont
  {S\'en\'echal}, \citenamefont {Perez},\ and\ \citenamefont
  {Pioro-Ladri\`ere}}]{Senechal2000}%
  \BibitemOpen
  \bibfield  {author} {\bibinfo {author} {\bibfnamefont {D.}~\bibnamefont
  {S\'en\'echal}}, \bibinfo {author} {\bibfnamefont {D.}~\bibnamefont {Perez}},
  \ and\ \bibinfo {author} {\bibfnamefont {M.}~\bibnamefont
  {Pioro-Ladri\`ere}},\ }\href {\doibase 10.1103/PhysRevLett.84.522} {\bibfield
   {journal} {\bibinfo  {journal} {Phys. Rev. Lett.}\ }\textbf {\bibinfo
  {volume} {84}},\ \bibinfo {pages} {522} (\bibinfo {year} {2000})}\BibitemShut
  {NoStop}%
\bibitem [{\citenamefont {S{\'{e}}n{\'{e}}chal}\ \emph
  {et~al.}(2002)\citenamefont {S{\'{e}}n{\'{e}}chal}, \citenamefont {Perez},\
  and\ \citenamefont {Plouffe}}]{Senechal2002}%
  \BibitemOpen
  \bibfield  {author} {\bibinfo {author} {\bibfnamefont {D.}~\bibnamefont
  {S{\'{e}}n{\'{e}}chal}}, \bibinfo {author} {\bibfnamefont {D.}~\bibnamefont
  {Perez}}, \ and\ \bibinfo {author} {\bibfnamefont {D.}~\bibnamefont
  {Plouffe}},\ }\href@noop {} {\bibfield  {journal} {\bibinfo  {journal} {Phys.
  Rev. B}\ }\textbf {\bibinfo {volume} {66}},\ \bibinfo {pages} {075129}
  (\bibinfo {year} {2002})}\BibitemShut {NoStop}%
\bibitem [{\citenamefont {Bulut}\ \emph {et~al.}(1994)\citenamefont {Bulut},
  \citenamefont {Scalapino},\ and\ \citenamefont {White}}]{Bulut1994}%
  \BibitemOpen
  \bibfield  {author} {\bibinfo {author} {\bibfnamefont {N.}~\bibnamefont
  {Bulut}}, \bibinfo {author} {\bibfnamefont {D.~J.}\ \bibnamefont
  {Scalapino}}, \ and\ \bibinfo {author} {\bibfnamefont {S.~R.}\ \bibnamefont
  {White}},\ }\href@noop {} {\bibfield  {journal} {\bibinfo  {journal} {Phys.
  Rev. Lett.}\ }\textbf {\bibinfo {volume} {72}},\ \bibinfo {pages} {705}
  (\bibinfo {year} {1994})}\BibitemShut {NoStop}%
\bibitem [{\citenamefont {Preuss}\ \emph {et~al.}(1995)\citenamefont {Preuss},
  \citenamefont {Hanke},\ and\ \citenamefont {von~der Linden}}]{Preuss1995}%
  \BibitemOpen
  \bibfield  {author} {\bibinfo {author} {\bibfnamefont {R.}~\bibnamefont
  {Preuss}}, \bibinfo {author} {\bibfnamefont {W.}~\bibnamefont {Hanke}}, \
  and\ \bibinfo {author} {\bibfnamefont {W.}~\bibnamefont {von~der Linden}},\
  }\href@noop {} {\bibfield  {journal} {\bibinfo  {journal} {Phys. Rev. Lett.}\
  }\textbf {\bibinfo {volume} {75}},\ \bibinfo {pages} {1344} (\bibinfo {year}
  {1995})}\BibitemShut {NoStop}%
\bibitem [{\citenamefont {Preuss}\ \emph {et~al.}(1997)\citenamefont {Preuss},
  \citenamefont {Hanke}, \citenamefont {Gr{\"{o}}ber},\ and\ \citenamefont
  {Evertz}}]{Preuss1997}%
  \BibitemOpen
  \bibfield  {author} {\bibinfo {author} {\bibfnamefont {R.}~\bibnamefont
  {Preuss}}, \bibinfo {author} {\bibfnamefont {W.}~\bibnamefont {Hanke}},
  \bibinfo {author} {\bibfnamefont {C.}~\bibnamefont {Gr{\"{o}}ber}}, \ and\
  \bibinfo {author} {\bibfnamefont {H.~G.}\ \bibnamefont {Evertz}},\
  }\href@noop {} {\bibfield  {journal} {\bibinfo  {journal} {Phys. Rev. Lett.}\
  }\textbf {\bibinfo {volume} {79}},\ \bibinfo {pages} {1122} (\bibinfo {year}
  {1997})}\BibitemShut {NoStop}%
\bibitem [{\citenamefont {Gr\"ober}\ \emph {et~al.}(2000)\citenamefont
  {Gr\"ober}, \citenamefont {Eder},\ and\ \citenamefont {Hanke}}]{Grober2000}%
  \BibitemOpen
  \bibfield  {author} {\bibinfo {author} {\bibfnamefont {C.}~\bibnamefont
  {Gr\"ober}}, \bibinfo {author} {\bibfnamefont {R.}~\bibnamefont {Eder}}, \
  and\ \bibinfo {author} {\bibfnamefont {W.}~\bibnamefont {Hanke}},\
  }\href@noop {} {\bibfield  {journal} {\bibinfo  {journal} {Phys. Rev. B}\
  }\textbf {\bibinfo {volume} {62}},\ \bibinfo {pages} {4336} (\bibinfo {year}
  {2000})}\BibitemShut {NoStop}%
\bibitem [{\citenamefont {Dahnken}\ \emph {et~al.}(2004)\citenamefont
  {Dahnken}, \citenamefont {Aichhorn}, \citenamefont {Hanke}, \citenamefont
  {Arrigoni},\ and\ \citenamefont {Potthoff}}]{Dahnken2004}%
  \BibitemOpen
  \bibfield  {author} {\bibinfo {author} {\bibfnamefont {C.}~\bibnamefont
  {Dahnken}}, \bibinfo {author} {\bibfnamefont {M.}~\bibnamefont {Aichhorn}},
  \bibinfo {author} {\bibfnamefont {W.}~\bibnamefont {Hanke}}, \bibinfo
  {author} {\bibfnamefont {E.}~\bibnamefont {Arrigoni}}, \ and\ \bibinfo
  {author} {\bibfnamefont {M.}~\bibnamefont {Potthoff}},\ }\href@noop {}
  {\bibfield  {journal} {\bibinfo  {journal} {Phys. Rev. B}\ }\textbf {\bibinfo
  {volume} {70}},\ \bibinfo {pages} {245110} (\bibinfo {year}
  {2004})}\BibitemShut {NoStop}%
\bibitem [{\citenamefont {Aichhorn}\ \emph {et~al.}(2006)\citenamefont
  {Aichhorn}, \citenamefont {Arrigoni}, \citenamefont {Potthoff},\ and\
  \citenamefont {Hanke}}]{Aichhorn2006}%
  \BibitemOpen
  \bibfield  {author} {\bibinfo {author} {\bibfnamefont {M.}~\bibnamefont
  {Aichhorn}}, \bibinfo {author} {\bibfnamefont {E.}~\bibnamefont {Arrigoni}},
  \bibinfo {author} {\bibfnamefont {M.}~\bibnamefont {Potthoff}}, \ and\
  \bibinfo {author} {\bibfnamefont {W.}~\bibnamefont {Hanke}},\ }\href
  {\doibase 10.1103/PhysRevB.74.235117} {\bibfield  {journal} {\bibinfo
  {journal} {Phys. Rev. B}\ }\textbf {\bibinfo {volume} {74}},\ \bibinfo
  {pages} {235117} (\bibinfo {year} {2006})}\BibitemShut {NoStop}%
\bibitem [{\citenamefont {Macridin}\ \emph {et~al.}(2006)\citenamefont
  {Macridin}, \citenamefont {Jarrell}, \citenamefont {Maier}, \citenamefont
  {Kent},\ and\ \citenamefont {D'azevedo}}]{Macridin2006}%
  \BibitemOpen
  \bibfield  {author} {\bibinfo {author} {\bibfnamefont {A.}~\bibnamefont
  {Macridin}}, \bibinfo {author} {\bibfnamefont {M.}~\bibnamefont {Jarrell}},
  \bibinfo {author} {\bibfnamefont {T.}~\bibnamefont {Maier}}, \bibinfo
  {author} {\bibfnamefont {P.~R.~C.}\ \bibnamefont {Kent}}, \ and\ \bibinfo
  {author} {\bibfnamefont {E.}~\bibnamefont {D'azevedo}},\ }\href@noop {}
  {\bibfield  {journal} {\bibinfo  {journal} {Phys. Rev. Lett.}\ }\textbf
  {\bibinfo {volume} {97}},\ \bibinfo {pages} {36401} (\bibinfo {year}
  {2006})}\BibitemShut {NoStop}%
\bibitem [{\citenamefont {Fabrizio}\ and\ \citenamefont
  {Parola}(1993)}]{Fabrizio1993a}%
  \BibitemOpen
  \bibfield  {author} {\bibinfo {author} {\bibfnamefont {M.}~\bibnamefont
  {Fabrizio}}\ and\ \bibinfo {author} {\bibfnamefont {A.}~\bibnamefont
  {Parola}},\ }\href {\doibase 10.1103/PhysRevLett.70.226} {\bibfield
  {journal} {\bibinfo  {journal} {Phys. Rev. Lett.}\ }\textbf {\bibinfo
  {volume} {70}},\ \bibinfo {pages} {226} (\bibinfo {year} {1993})}\BibitemShut
  {NoStop}%
\bibitem [{\citenamefont {Fabrizio}(1993)}]{Fabrizio1993b}%
  \BibitemOpen
  \bibfield  {author} {\bibinfo {author} {\bibfnamefont {M.}~\bibnamefont
  {Fabrizio}},\ }\href {\doibase 10.1103/PhysRevB.48.15838} {\bibfield
  {journal} {\bibinfo  {journal} {Phys. Rev. B}\ }\textbf {\bibinfo {volume}
  {48}},\ \bibinfo {pages} {15838} (\bibinfo {year} {1993})}\BibitemShut
  {NoStop}%
\bibitem [{\citenamefont {Castellani}\ \emph {et~al.}(1994)\citenamefont
  {Castellani}, \citenamefont {Di~Castro},\ and\ \citenamefont
  {Metzner}}]{Castellani1994}%
  \BibitemOpen
  \bibfield  {author} {\bibinfo {author} {\bibfnamefont {C.}~\bibnamefont
  {Castellani}}, \bibinfo {author} {\bibfnamefont {C.}~\bibnamefont
  {Di~Castro}}, \ and\ \bibinfo {author} {\bibfnamefont {W.}~\bibnamefont
  {Metzner}},\ }\href {\doibase 10.1103/PhysRevLett.72.316} {\bibfield
  {journal} {\bibinfo  {journal} {Phys. Rev. Lett.}\ }\textbf {\bibinfo
  {volume} {72}},\ \bibinfo {pages} {316} (\bibinfo {year} {1994})}\BibitemShut
  {NoStop}%
\bibitem [{\citenamefont {Balents}\ and\ \citenamefont
  {Fisher}(1996)}]{Balents1996}%
  \BibitemOpen
  \bibfield  {author} {\bibinfo {author} {\bibfnamefont {L.}~\bibnamefont
  {Balents}}\ and\ \bibinfo {author} {\bibfnamefont {M.~P.~A.}\ \bibnamefont
  {Fisher}},\ }\href@noop {} {\bibfield  {journal} {\bibinfo  {journal} {Phys.
  Rev. B}\ }\textbf {\bibinfo {volume} {53}},\ \bibinfo {pages} {12133}
  (\bibinfo {year} {1996})}\BibitemShut {NoStop}%
\bibitem [{\citenamefont {Noack}\ \emph {et~al.}(1994)\citenamefont {Noack},
  \citenamefont {White},\ and\ \citenamefont {Scalapino}}]{Noack1994}%
  \BibitemOpen
  \bibfield  {author} {\bibinfo {author} {\bibfnamefont {R.~M.}\ \bibnamefont
  {Noack}}, \bibinfo {author} {\bibfnamefont {S.~R.}\ \bibnamefont {White}}, \
  and\ \bibinfo {author} {\bibfnamefont {D.~J.}\ \bibnamefont {Scalapino}},\
  }\href@noop {} {\bibfield  {journal} {\bibinfo  {journal} {Phys. Rev. Lett.}\
  }\textbf {\bibinfo {volume} {73}},\ \bibinfo {pages} {882} (\bibinfo {year}
  {1994})}\BibitemShut {NoStop}%
\bibitem [{\citenamefont {Noack}\ \emph {et~al.}(1996)\citenamefont {Noack},
  \citenamefont {White},\ and\ \citenamefont {Scalapino}}]{Noack1996}%
  \BibitemOpen
  \bibfield  {author} {\bibinfo {author} {\bibfnamefont {R.}~\bibnamefont
  {Noack}}, \bibinfo {author} {\bibfnamefont {S.}~\bibnamefont {White}}, \ and\
  \bibinfo {author} {\bibfnamefont {D.}~\bibnamefont {Scalapino}},\ }\href
  {\doibase https://doi.org/10.1016/S0921-4534(96)00515-1} {\bibfield
  {journal} {\bibinfo  {journal} {Physica C: Superconductivity}\ }\textbf
  {\bibinfo {volume} {270}},\ \bibinfo {pages} {281 } (\bibinfo {year}
  {1996})}\BibitemShut {NoStop}%
\bibitem [{\citenamefont {Lin}\ \emph {et~al.}(1998)\citenamefont {Lin},
  \citenamefont {Balents},\ and\ \citenamefont {Fisher}}]{Lin1998}%
  \BibitemOpen
  \bibfield  {author} {\bibinfo {author} {\bibfnamefont {H.-H.}\ \bibnamefont
  {Lin}}, \bibinfo {author} {\bibfnamefont {L.}~\bibnamefont {Balents}}, \ and\
  \bibinfo {author} {\bibfnamefont {M.~P.~A.}\ \bibnamefont {Fisher}},\ }\href
  {\doibase 10.1103/PhysRevB.58.1794} {\bibfield  {journal} {\bibinfo
  {journal} {Phys. Rev. B}\ }\textbf {\bibinfo {volume} {58}},\ \bibinfo
  {pages} {1794} (\bibinfo {year} {1998})}\BibitemShut {NoStop}%
\bibitem [{\citenamefont {Assaraf}\ \emph {et~al.}(2004)\citenamefont
  {Assaraf}, \citenamefont {Azaria}, \citenamefont {Boulat}, \citenamefont
  {Caffarel},\ and\ \citenamefont {Lecheminant}}]{Assaraf2004}%
  \BibitemOpen
  \bibfield  {author} {\bibinfo {author} {\bibfnamefont {R.}~\bibnamefont
  {Assaraf}}, \bibinfo {author} {\bibfnamefont {P.}~\bibnamefont {Azaria}},
  \bibinfo {author} {\bibfnamefont {E.}~\bibnamefont {Boulat}}, \bibinfo
  {author} {\bibfnamefont {M.}~\bibnamefont {Caffarel}}, \ and\ \bibinfo
  {author} {\bibfnamefont {P.}~\bibnamefont {Lecheminant}},\ }\href {\doibase
  10.1103/PhysRevLett.93.016407} {\bibfield  {journal} {\bibinfo  {journal}
  {Phys. Rev. Lett.}\ }\textbf {\bibinfo {volume} {93}},\ \bibinfo {pages}
  {016407} (\bibinfo {year} {2004})}\BibitemShut {NoStop}%
\bibitem [{\citenamefont {Dagotto}(1994)}]{Dagotto1994b}%
  \BibitemOpen
  \bibfield  {author} {\bibinfo {author} {\bibfnamefont {E.}~\bibnamefont
  {Dagotto}},\ }\href@noop {} {\bibfield  {journal} {\bibinfo  {journal} {Rev.
  Mod. Phys.}\ }\textbf {\bibinfo {volume} {66}},\ \bibinfo {pages} {763}
  (\bibinfo {year} {1994})}\BibitemShut {NoStop}%
\bibitem [{\citenamefont {Chernyshev}\ and\ \citenamefont
  {Leung}(1999)}]{Chernyshev1999}%
  \BibitemOpen
  \bibfield  {author} {\bibinfo {author} {\bibfnamefont {A.~L.}\ \bibnamefont
  {Chernyshev}}\ and\ \bibinfo {author} {\bibfnamefont {P.~W.}\ \bibnamefont
  {Leung}},\ }\href {\doibase 10.1103/PhysRevB.60.1592} {\bibfield  {journal}
  {\bibinfo  {journal} {Phys. Rev. B}\ }\textbf {\bibinfo {volume} {60}},\
  \bibinfo {pages} {1592} (\bibinfo {year} {1999})}\BibitemShut {NoStop}%
\bibitem [{\citenamefont {Chernyshev}\ \emph {et~al.}(2000)\citenamefont
  {Chernyshev}, \citenamefont {Castro~Neto},\ and\ \citenamefont
  {Bishop}}]{Chernyshev2000}%
  \BibitemOpen
  \bibfield  {author} {\bibinfo {author} {\bibfnamefont {A.~L.}\ \bibnamefont
  {Chernyshev}}, \bibinfo {author} {\bibfnamefont {A.~H.}\ \bibnamefont
  {Castro~Neto}}, \ and\ \bibinfo {author} {\bibfnamefont {A.~R.}\ \bibnamefont
  {Bishop}},\ }\href {\doibase 10.1103/PhysRevLett.84.4922} {\bibfield
  {journal} {\bibinfo  {journal} {Phys. Rev. Lett.}\ }\textbf {\bibinfo
  {volume} {84}},\ \bibinfo {pages} {4922} (\bibinfo {year}
  {2000})}\BibitemShut {NoStop}%
\bibitem [{\citenamefont {Chernyshev}\ \emph {et~al.}(2002)\citenamefont
  {Chernyshev}, \citenamefont {White},\ and\ \citenamefont
  {Castro~Neto}}]{Chernyshev2002}%
  \BibitemOpen
  \bibfield  {author} {\bibinfo {author} {\bibfnamefont {A.~L.}\ \bibnamefont
  {Chernyshev}}, \bibinfo {author} {\bibfnamefont {S.~R.}\ \bibnamefont
  {White}}, \ and\ \bibinfo {author} {\bibfnamefont {A.~H.}\ \bibnamefont
  {Castro~Neto}},\ }\href {\doibase 10.1103/PhysRevB.65.214527} {\bibfield
  {journal} {\bibinfo  {journal} {Phys. Rev. B}\ }\textbf {\bibinfo {volume}
  {65}},\ \bibinfo {pages} {214527} (\bibinfo {year} {2002})}\BibitemShut
  {NoStop}%
\bibitem [{\citenamefont {Chernyshev}\ and\ \citenamefont
  {Wood}(2003)}]{Chernyshev2003}%
  \BibitemOpen
  \bibfield  {author} {\bibinfo {author} {\bibfnamefont {A.~L.}\ \bibnamefont
  {Chernyshev}}\ and\ \bibinfo {author} {\bibfnamefont {R.~F.}\ \bibnamefont
  {Wood}},\ }in\ \href@noop {} {\emph {\bibinfo {booktitle} {Models and Methods
  of High-Tc Superconductivity: Some Frontal Aspects}}},\ Vol.~\bibinfo
  {volume} {1},\ \bibinfo {editor} {edited by\ \bibinfo {editor} {\bibfnamefont
  {J.~K.}\ \bibnamefont {Srivastava}}\ and\ \bibinfo {editor} {\bibfnamefont
  {S.~M.}\ \bibnamefont {Rao}}}\ (\bibinfo  {publisher} {Nova Science
  Publishers, Inc., Hauppauge NY},\ \bibinfo {year} {2003})\ Chap.~\bibinfo
  {chapter} {11}\BibitemShut {NoStop}%
\bibitem [{\citenamefont {\ifmmode~\check{S}\else \v{S}\fi{}makov}\ \emph
  {et~al.}(2007{\natexlab{a}})\citenamefont {\ifmmode~\check{S}\else
  \v{S}\fi{}makov}, \citenamefont {Chernyshev},\ and\ \citenamefont
  {White}}]{Smakov2007}%
  \BibitemOpen
  \bibfield  {author} {\bibinfo {author} {\bibfnamefont {J.}~\bibnamefont
  {\ifmmode~\check{S}\else \v{S}\fi{}makov}}, \bibinfo {author} {\bibfnamefont
  {A.~L.}\ \bibnamefont {Chernyshev}}, \ and\ \bibinfo {author} {\bibfnamefont
  {S.~R.}\ \bibnamefont {White}},\ }\href {\doibase
  10.1103/PhysRevLett.98.266401} {\bibfield  {journal} {\bibinfo  {journal}
  {Phys. Rev. Lett.}\ }\textbf {\bibinfo {volume} {98}},\ \bibinfo {pages}
  {266401} (\bibinfo {year} {2007}{\natexlab{a}})}\BibitemShut {NoStop}%
\bibitem [{\citenamefont {\ifmmode~\check{S}\else \v{S}\fi{}makov}\ \emph
  {et~al.}(2007{\natexlab{b}})\citenamefont {\ifmmode~\check{S}\else
  \v{S}\fi{}makov}, \citenamefont {Chernyshev},\ and\ \citenamefont
  {White}}]{Smakov2007b}%
  \BibitemOpen
  \bibfield  {author} {\bibinfo {author} {\bibfnamefont {J.}~\bibnamefont
  {\ifmmode~\check{S}\else \v{S}\fi{}makov}}, \bibinfo {author} {\bibfnamefont
  {A.~L.}\ \bibnamefont {Chernyshev}}, \ and\ \bibinfo {author} {\bibfnamefont
  {S.~R.}\ \bibnamefont {White}},\ }\href {\doibase 10.1103/PhysRevB.76.115106}
  {\bibfield  {journal} {\bibinfo  {journal} {Phys. Rev. B}\ }\textbf {\bibinfo
  {volume} {76}},\ \bibinfo {pages} {115106} (\bibinfo {year}
  {2007}{\natexlab{b}})}\BibitemShut {NoStop}%
\bibitem [{\citenamefont {Eder}(1998)}]{Eder1998}%
  \BibitemOpen
  \bibfield  {author} {\bibinfo {author} {\bibfnamefont {R.}~\bibnamefont
  {Eder}},\ }\href {\doibase 10.1103/PhysRevB.57.12832} {\bibfield  {journal}
  {\bibinfo  {journal} {Phys. Rev. B}\ }\textbf {\bibinfo {volume} {57}},\
  \bibinfo {pages} {12832} (\bibinfo {year} {1998})}\BibitemShut {NoStop}%
\bibitem [{\citenamefont {Sushkov}(1999)}]{Sushkov1999}%
  \BibitemOpen
  \bibfield  {author} {\bibinfo {author} {\bibfnamefont {O.~P.}\ \bibnamefont
  {Sushkov}},\ }\href {\doibase 10.1103/PhysRevB.60.3289} {\bibfield  {journal}
  {\bibinfo  {journal} {Phys. Rev. B}\ }\textbf {\bibinfo {volume} {60}},\
  \bibinfo {pages} {3289} (\bibinfo {year} {1999})}\BibitemShut {NoStop}%
\bibitem [{\citenamefont {Endres}\ \emph {et~al.}(1996)\citenamefont {Endres},
  \citenamefont {Noack}, \citenamefont {Hanke}, \citenamefont {Poilblanc},\
  and\ \citenamefont {Scalapino}}]{Endres1996}%
  \BibitemOpen
  \bibfield  {author} {\bibinfo {author} {\bibfnamefont {H.}~\bibnamefont
  {Endres}}, \bibinfo {author} {\bibfnamefont {R.~M.}\ \bibnamefont {Noack}},
  \bibinfo {author} {\bibfnamefont {W.}~\bibnamefont {Hanke}}, \bibinfo
  {author} {\bibfnamefont {D.}~\bibnamefont {Poilblanc}}, \ and\ \bibinfo
  {author} {\bibfnamefont {D.~J.}\ \bibnamefont {Scalapino}},\ }\href {\doibase
  10.1103/PhysRevB.53.5530} {\bibfield  {journal} {\bibinfo  {journal} {Phys.
  Rev. B}\ }\textbf {\bibinfo {volume} {53}},\ \bibinfo {pages} {5530}
  (\bibinfo {year} {1996})}\BibitemShut {NoStop}%
\bibitem [{\citenamefont {Brunner}\ \emph {et~al.}(2001)\citenamefont
  {Brunner}, \citenamefont {Capponi}, \citenamefont {Assaad},\ and\
  \citenamefont {Muramatsu}}]{Brunner2001}%
  \BibitemOpen
  \bibfield  {author} {\bibinfo {author} {\bibfnamefont {M.}~\bibnamefont
  {Brunner}}, \bibinfo {author} {\bibfnamefont {S.}~\bibnamefont {Capponi}},
  \bibinfo {author} {\bibfnamefont {F.~F.}\ \bibnamefont {Assaad}}, \ and\
  \bibinfo {author} {\bibfnamefont {A.}~\bibnamefont {Muramatsu}},\ }\href@noop
  {} {\bibfield  {journal} {\bibinfo  {journal} {Phys. Rev. B}\ }\textbf
  {\bibinfo {volume} {63}},\ \bibinfo {pages} {180511} (\bibinfo {year}
  {2001})}\BibitemShut {NoStop}%
\bibitem [{\citenamefont {Poilblanc}\ \emph {et~al.}(1995)\citenamefont
  {Poilblanc}, \citenamefont {Scalapino},\ and\ \citenamefont
  {Hanke}}]{Poilblanc1995}%
  \BibitemOpen
  \bibfield  {author} {\bibinfo {author} {\bibfnamefont {D.}~\bibnamefont
  {Poilblanc}}, \bibinfo {author} {\bibfnamefont {D.~J.}\ \bibnamefont
  {Scalapino}}, \ and\ \bibinfo {author} {\bibfnamefont {W.}~\bibnamefont
  {Hanke}},\ }\href@noop {} {\bibfield  {journal} {\bibinfo  {journal} {Phys.
  Rev. B}\ }\textbf {\bibinfo {volume} {52}},\ \bibinfo {pages} {6796}
  (\bibinfo {year} {1995})}\BibitemShut {NoStop}%
\bibitem [{\citenamefont {Haas}\ and\ \citenamefont
  {Dagotto}(1996)}]{Haas1996a}%
  \BibitemOpen
  \bibfield  {author} {\bibinfo {author} {\bibfnamefont {S.}~\bibnamefont
  {Haas}}\ and\ \bibinfo {author} {\bibfnamefont {E.}~\bibnamefont {Dagotto}},\
  }\href {\doibase 10.1103/PhysRevB.54.R3718} {\bibfield  {journal} {\bibinfo
  {journal} {Phys. Rev. B}\ }\textbf {\bibinfo {volume} {54}},\ \bibinfo
  {pages} {R3718} (\bibinfo {year} {1996})}\BibitemShut {NoStop}%
\bibitem [{\citenamefont {Rice}\ \emph {et~al.}(1997)\citenamefont {Rice},
  \citenamefont {Haas}, \citenamefont {Sigrist},\ and\ \citenamefont
  {Zhang}}]{Rice1997}%
  \BibitemOpen
  \bibfield  {author} {\bibinfo {author} {\bibfnamefont {T.~M.}\ \bibnamefont
  {Rice}}, \bibinfo {author} {\bibfnamefont {S.}~\bibnamefont {Haas}}, \bibinfo
  {author} {\bibfnamefont {M.}~\bibnamefont {Sigrist}}, \ and\ \bibinfo
  {author} {\bibfnamefont {F.-C.}\ \bibnamefont {Zhang}},\ }\href {\doibase
  10.1103/PhysRevB.56.14655} {\bibfield  {journal} {\bibinfo  {journal} {Phys.
  Rev. B}\ }\textbf {\bibinfo {volume} {56}},\ \bibinfo {pages} {14655}
  (\bibinfo {year} {1997})}\BibitemShut {NoStop}%
\bibitem [{\citenamefont {Martins}\ \emph
  {et~al.}(2000{\natexlab{a}})\citenamefont {Martins}, \citenamefont {Gazza},
  \citenamefont {Xavier}, \citenamefont {Feiguin},\ and\ \citenamefont
  {Dagotto}}]{Martins2000}%
  \BibitemOpen
  \bibfield  {author} {\bibinfo {author} {\bibfnamefont {G.~B.}\ \bibnamefont
  {Martins}}, \bibinfo {author} {\bibfnamefont {C.}~\bibnamefont {Gazza}},
  \bibinfo {author} {\bibfnamefont {J.~C.}\ \bibnamefont {Xavier}}, \bibinfo
  {author} {\bibfnamefont {A.}~\bibnamefont {Feiguin}}, \ and\ \bibinfo
  {author} {\bibfnamefont {E.}~\bibnamefont {Dagotto}},\ }\href@noop {}
  {\bibfield  {journal} {\bibinfo  {journal} {Phys. Rev. Lett.}\ }\textbf
  {\bibinfo {volume} {84}},\ \bibinfo {pages} {5844} (\bibinfo {year}
  {2000}{\natexlab{a}})}\BibitemShut {NoStop}%
\bibitem [{\citenamefont {Zhu}\ \emph {et~al.}(2013)\citenamefont {Zhu},
  \citenamefont {Jiang}, \citenamefont {Qi}, \citenamefont {Tian},\ and\
  \citenamefont {Weng}}]{Zhu2013}%
  \BibitemOpen
  \bibfield  {author} {\bibinfo {author} {\bibfnamefont {Z.}~\bibnamefont
  {Zhu}}, \bibinfo {author} {\bibfnamefont {H.-C.}\ \bibnamefont {Jiang}},
  \bibinfo {author} {\bibfnamefont {Y.}~\bibnamefont {Qi}}, \bibinfo {author}
  {\bibfnamefont {C.}~\bibnamefont {Tian}}, \ and\ \bibinfo {author}
  {\bibfnamefont {Z.-Y.}\ \bibnamefont {Weng}},\ }\href
  {http://dx.doi.org/10.1038/srep02586} {\bibfield  {journal} {\bibinfo
  {journal} {Scientific Reports}\ }\textbf {\bibinfo {volume} {3}},\ \bibinfo
  {pages} {2586} (\bibinfo {year} {2013})},\ \bibinfo {note}
  {article}\BibitemShut {NoStop}%
\bibitem [{\citenamefont {Zhu}\ \emph {et~al.}(2014)\citenamefont {Zhu},
  \citenamefont {Jiang}, \citenamefont {Sheng},\ and\ \citenamefont
  {Weng}}]{Zhu2014}%
  \BibitemOpen
  \bibfield  {author} {\bibinfo {author} {\bibfnamefont {Z.}~\bibnamefont
  {Zhu}}, \bibinfo {author} {\bibfnamefont {H.-C.}\ \bibnamefont {Jiang}},
  \bibinfo {author} {\bibfnamefont {D.~N.}\ \bibnamefont {Sheng}}, \ and\
  \bibinfo {author} {\bibfnamefont {Z.-Y.}\ \bibnamefont {Weng}},\ }\href
  {http://dx.doi.org/10.1038/srep05419} {\bibfield  {journal} {\bibinfo
  {journal} {Scientific Reports}\ }\textbf {\bibinfo {volume} {4}},\ \bibinfo
  {pages} {5419} (\bibinfo {year} {2014})},\ \bibinfo {note}
  {article}\BibitemShut {NoStop}%
\bibitem [{\citenamefont {Zhu}\ \emph {et~al.}(2015)\citenamefont {Zhu},
  \citenamefont {Tian}, \citenamefont {Jiang}, \citenamefont {Qi},
  \citenamefont {Weng},\ and\ \citenamefont {Zaanen}}]{Zhu2015}%
  \BibitemOpen
  \bibfield  {author} {\bibinfo {author} {\bibfnamefont {Z.}~\bibnamefont
  {Zhu}}, \bibinfo {author} {\bibfnamefont {C.}~\bibnamefont {Tian}}, \bibinfo
  {author} {\bibfnamefont {H.-C.}\ \bibnamefont {Jiang}}, \bibinfo {author}
  {\bibfnamefont {Y.}~\bibnamefont {Qi}}, \bibinfo {author} {\bibfnamefont
  {Z.-Y.}\ \bibnamefont {Weng}}, \ and\ \bibinfo {author} {\bibfnamefont
  {J.}~\bibnamefont {Zaanen}},\ }\href {\doibase 10.1103/PhysRevB.92.035113}
  {\bibfield  {journal} {\bibinfo  {journal} {Phys. Rev. B}\ }\textbf {\bibinfo
  {volume} {92}},\ \bibinfo {pages} {035113} (\bibinfo {year}
  {2015})}\BibitemShut {NoStop}%
\bibitem [{\citenamefont {Zhu}\ and\ \citenamefont {Weng}(2015)}]{Zhu2015b}%
  \BibitemOpen
  \bibfield  {author} {\bibinfo {author} {\bibfnamefont {Z.}~\bibnamefont
  {Zhu}}\ and\ \bibinfo {author} {\bibfnamefont {Z.-Y.}\ \bibnamefont {Weng}},\
  }\href {\doibase 10.1103/PhysRevB.92.235156} {\bibfield  {journal} {\bibinfo
  {journal} {Phys. Rev. B}\ }\textbf {\bibinfo {volume} {92}},\ \bibinfo
  {pages} {235156} (\bibinfo {year} {2015})}\BibitemShut {NoStop}%
\bibitem [{\citenamefont {Zhu}\ \emph {et~al.}(2016)\citenamefont {Zhu},
  \citenamefont {Wang}, \citenamefont {Sheng},\ and\ \citenamefont
  {Weng}}]{Zhu2016}%
  \BibitemOpen
  \bibfield  {author} {\bibinfo {author} {\bibfnamefont {Z.}~\bibnamefont
  {Zhu}}, \bibinfo {author} {\bibfnamefont {Q.-R.}\ \bibnamefont {Wang}},
  \bibinfo {author} {\bibfnamefont {D.}~\bibnamefont {Sheng}}, \ and\ \bibinfo
  {author} {\bibfnamefont {Z.-Y.}\ \bibnamefont {Weng}},\ }\href {\doibase
  https://doi.org/10.1016/j.nuclphysb.2015.12.004} {\bibfield  {journal}
  {\bibinfo  {journal} {Nuclear Physics B}\ }\textbf {\bibinfo {volume}
  {903}},\ \bibinfo {pages} {51 } (\bibinfo {year} {2016})}\BibitemShut
  {NoStop}%
\bibitem [{\citenamefont {Zhu}\ \emph {et~al.}(2018)\citenamefont {Zhu},
  \citenamefont {Sheng},\ and\ \citenamefont {Weng}}]{Zhu2018}%
  \BibitemOpen
  \bibfield  {author} {\bibinfo {author} {\bibfnamefont {Z.}~\bibnamefont
  {Zhu}}, \bibinfo {author} {\bibfnamefont {D.~N.}\ \bibnamefont {Sheng}}, \
  and\ \bibinfo {author} {\bibfnamefont {Z.-Y.}\ \bibnamefont {Weng}},\ }\href
  {\doibase 10.1103/PhysRevB.98.035129} {\bibfield  {journal} {\bibinfo
  {journal} {Phys. Rev. B}\ }\textbf {\bibinfo {volume} {98}},\ \bibinfo
  {pages} {035129} (\bibinfo {year} {2018})}\BibitemShut {NoStop}%
\bibitem [{\citenamefont {White}\ \emph {et~al.}(2015)\citenamefont {White},
  \citenamefont {Scalapino},\ and\ \citenamefont {Kivelson}}]{White2015}%
  \BibitemOpen
  \bibfield  {author} {\bibinfo {author} {\bibfnamefont {S.~R.}\ \bibnamefont
  {White}}, \bibinfo {author} {\bibfnamefont {D.~J.}\ \bibnamefont
  {Scalapino}}, \ and\ \bibinfo {author} {\bibfnamefont {S.~A.}\ \bibnamefont
  {Kivelson}},\ }\href {\doibase 10.1103/PhysRevLett.115.056401} {\bibfield
  {journal} {\bibinfo  {journal} {Phys. Rev. Lett.}\ }\textbf {\bibinfo
  {volume} {115}},\ \bibinfo {pages} {056401} (\bibinfo {year}
  {2015})}\BibitemShut {NoStop}%
\bibitem [{\citenamefont {Troyer}\ \emph {et~al.}(1996)\citenamefont {Troyer},
  \citenamefont {Tsunetsugu},\ and\ \citenamefont {Rice}}]{Troyer1996}%
  \BibitemOpen
  \bibfield  {author} {\bibinfo {author} {\bibfnamefont {M.}~\bibnamefont
  {Troyer}}, \bibinfo {author} {\bibfnamefont {H.}~\bibnamefont {Tsunetsugu}},
  \ and\ \bibinfo {author} {\bibfnamefont {T.~M.}\ \bibnamefont {Rice}},\
  }\href {\doibase 10.1103/PhysRevB.53.251} {\bibfield  {journal} {\bibinfo
  {journal} {Phys. Rev. B}\ }\textbf {\bibinfo {volume} {53}},\ \bibinfo
  {pages} {251} (\bibinfo {year} {1996})}\BibitemShut {NoStop}%
\bibitem [{\citenamefont {Riera}\ \emph {et~al.}(1999)\citenamefont {Riera},
  \citenamefont {Poilblanc},\ and\ \citenamefont {Dagotto}}]{Riera1999}%
  \BibitemOpen
  \bibfield  {author} {\bibinfo {author} {\bibfnamefont {J.}~\bibnamefont
  {Riera}}, \bibinfo {author} {\bibfnamefont {D.}~\bibnamefont {Poilblanc}}, \
  and\ \bibinfo {author} {\bibfnamefont {E.}~\bibnamefont {Dagotto}},\ }\href
  {\doibase 10.1007/s100510050588} {\bibfield  {journal} {\bibinfo  {journal}
  {The European Physical Journal B - Condensed Matter and Complex Systems}\
  }\textbf {\bibinfo {volume} {7}},\ \bibinfo {pages} {53} (\bibinfo {year}
  {1999})}\BibitemShut {NoStop}%
\bibitem [{\citenamefont {Liu}\ \emph {et~al.}(2016)\citenamefont {Liu},
  \citenamefont {Jiang},\ and\ \citenamefont {Devereaux}}]{Liu2016}%
  \BibitemOpen
  \bibfield  {author} {\bibinfo {author} {\bibfnamefont {S.}~\bibnamefont
  {Liu}}, \bibinfo {author} {\bibfnamefont {H.-C.}\ \bibnamefont {Jiang}}, \
  and\ \bibinfo {author} {\bibfnamefont {T.~P.}\ \bibnamefont {Devereaux}},\
  }\href {\doibase 10.1103/PhysRevB.94.155149} {\bibfield  {journal} {\bibinfo
  {journal} {Phys. Rev. B}\ }\textbf {\bibinfo {volume} {94}},\ \bibinfo
  {pages} {155149} (\bibinfo {year} {2016})}\BibitemShut {NoStop}%
\bibitem [{\citenamefont {Barthel}\ \emph {et~al.}(2009)\citenamefont
  {Barthel}, \citenamefont {Schollw\"ock},\ and\ \citenamefont
  {White}}]{Barthel2009}%
  \BibitemOpen
  \bibfield  {author} {\bibinfo {author} {\bibfnamefont {T.}~\bibnamefont
  {Barthel}}, \bibinfo {author} {\bibfnamefont {U.}~\bibnamefont
  {Schollw\"ock}}, \ and\ \bibinfo {author} {\bibfnamefont {S.~R.}\
  \bibnamefont {White}},\ }\href@noop {} {\bibfield  {journal} {\bibinfo
  {journal} {Phys. Rev. B}\ }\textbf {\bibinfo {volume} {79}},\ \bibinfo
  {pages} {245101} (\bibinfo {year} {2009})}\BibitemShut {NoStop}%
\bibitem [{\citenamefont {Fiete}\ and\ \citenamefont
  {Balents}(2004)}]{Fiete2004}%
  \BibitemOpen
  \bibfield  {author} {\bibinfo {author} {\bibfnamefont {G.~A.}\ \bibnamefont
  {Fiete}}\ and\ \bibinfo {author} {\bibfnamefont {L.}~\bibnamefont
  {Balents}},\ }\href@noop {} {\bibfield  {journal} {\bibinfo  {journal} {Phys.
  Rev. Lett.}\ }\textbf {\bibinfo {volume} {93}},\ \bibinfo {pages} {226401}
  (\bibinfo {year} {2004})}\BibitemShut {NoStop}%
\bibitem [{\citenamefont {Cheianov}\ and\ \citenamefont
  {Zvonarev}(2004)}]{Cheianov2004}%
  \BibitemOpen
  \bibfield  {author} {\bibinfo {author} {\bibfnamefont {V.~V.}\ \bibnamefont
  {Cheianov}}\ and\ \bibinfo {author} {\bibfnamefont {M.~B.}\ \bibnamefont
  {Zvonarev}},\ }\href@noop {} {\bibfield  {journal} {\bibinfo  {journal} {Phys
  Rev Lett}\ }\textbf {\bibinfo {volume} {92}},\ \bibinfo {pages} {176401}
  (\bibinfo {year} {2004})}\BibitemShut {NoStop}%
\bibitem [{\citenamefont {Cheianov}\ \emph {et~al.}(2005)\citenamefont
  {Cheianov}, \citenamefont {Smith},\ and\ \citenamefont
  {Zvoranev}}]{Cheianov2005}%
  \BibitemOpen
  \bibfield  {author} {\bibinfo {author} {\bibfnamefont {V.~V.}\ \bibnamefont
  {Cheianov}}, \bibinfo {author} {\bibfnamefont {H.}~\bibnamefont {Smith}}, \
  and\ \bibinfo {author} {\bibfnamefont {M.~B.}\ \bibnamefont {Zvoranev}},\
  }\href@noop {} {\bibfield  {journal} {\bibinfo  {journal} {Phys. Rev. A}\
  }\textbf {\bibinfo {volume} {71}},\ \bibinfo {pages} {033610} (\bibinfo
  {year} {2005})}\BibitemShut {NoStop}%
\bibitem [{\citenamefont {Fiete}(2006)}]{Fiete2006}%
  \BibitemOpen
  \bibfield  {author} {\bibinfo {author} {\bibfnamefont {G.~A.}\ \bibnamefont
  {Fiete}},\ }\href@noop {} {\bibfield  {journal} {\bibinfo  {journal} {Phys.
  Rev. Lett.}\ }\textbf {\bibinfo {volume} {97}},\ \bibinfo {pages} {256403}
  (\bibinfo {year} {2006})}\BibitemShut {NoStop}%
\bibitem [{\citenamefont {Fiete}(2007)}]{Fiete2007b}%
  \BibitemOpen
  \bibfield  {author} {\bibinfo {author} {\bibfnamefont {G.}~\bibnamefont
  {Fiete}},\ }\href@noop {} {\bibfield  {journal} {\bibinfo  {journal} {Rev.
  Mod. Phys.}\ }\textbf {\bibinfo {volume} {79}},\ \bibinfo {pages} {801}
  (\bibinfo {year} {2007})}\BibitemShut {NoStop}%
\bibitem [{\citenamefont {Halperin}(2007)}]{Halperin2007}%
  \BibitemOpen
  \bibfield  {author} {\bibinfo {author} {\bibfnamefont {B.~I.}\ \bibnamefont
  {Halperin}},\ }\href@noop {} {\bibfield  {journal} {\bibinfo  {journal} {J.
  Appl. Phys.}\ }\textbf {\bibinfo {volume} {101}},\ \bibinfo {pages} {081601}
  (\bibinfo {year} {2007})}\BibitemShut {NoStop}%
\bibitem [{\citenamefont {Feiguin}\ and\ \citenamefont
  {Fiete}(2010)}]{Feiguin2009d}%
  \BibitemOpen
  \bibfield  {author} {\bibinfo {author} {\bibfnamefont {A.~E.}\ \bibnamefont
  {Feiguin}}\ and\ \bibinfo {author} {\bibfnamefont {G.~A.}\ \bibnamefont
  {Fiete}},\ }\href@noop {} {\bibfield  {journal} {\bibinfo  {journal} {Phys.
  Rev. B}\ }\textbf {\bibinfo {volume} {81}},\ \bibinfo {pages} {075108}
  (\bibinfo {year} {2010})}\BibitemShut {NoStop}%
\bibitem [{\citenamefont {Nocera}\ \emph {et~al.}(2018)\citenamefont {Nocera},
  \citenamefont {Essler},\ and\ \citenamefont {Feiguin}}]{Nocera2018}%
  \BibitemOpen
  \bibfield  {author} {\bibinfo {author} {\bibfnamefont {A.}~\bibnamefont
  {Nocera}}, \bibinfo {author} {\bibfnamefont {F.~H.~L.}\ \bibnamefont
  {Essler}}, \ and\ \bibinfo {author} {\bibfnamefont {A.~E.}\ \bibnamefont
  {Feiguin}},\ }\href {\doibase 10.1103/PhysRevB.97.045146} {\bibfield
  {journal} {\bibinfo  {journal} {Phys. Rev. B}\ }\textbf {\bibinfo {volume}
  {97}},\ \bibinfo {pages} {045146} (\bibinfo {year} {2018})}\BibitemShut
  {NoStop}%
\bibitem [{\citenamefont {Essler}\ and\ \citenamefont
  {Tsvelik}(2002)}]{Essler2002}%
  \BibitemOpen
  \bibfield  {author} {\bibinfo {author} {\bibfnamefont {F.~H.~L.}\
  \bibnamefont {Essler}}\ and\ \bibinfo {author} {\bibfnamefont {A.~M.}\
  \bibnamefont {Tsvelik}},\ }\href {\doibase 10.1103/PhysRevB.65.115117}
  {\bibfield  {journal} {\bibinfo  {journal} {Phys. Rev. B}\ }\textbf {\bibinfo
  {volume} {65}},\ \bibinfo {pages} {115117} (\bibinfo {year}
  {2002})}\BibitemShut {NoStop}%
\bibitem [{\citenamefont {Schrieffer}\ \emph {et~al.}(1988)\citenamefont
  {Schrieffer}, \citenamefont {Wen},\ and\ \citenamefont
  {Zhang}}]{Schrieffer1988}%
  \BibitemOpen
  \bibfield  {author} {\bibinfo {author} {\bibfnamefont {J.~R.}\ \bibnamefont
  {Schrieffer}}, \bibinfo {author} {\bibfnamefont {X.-G.}\ \bibnamefont {Wen}},
  \ and\ \bibinfo {author} {\bibfnamefont {Z.-C.}\ \bibnamefont {Zhang}},\
  }\href@noop {} {\bibfield  {journal} {\bibinfo  {journal} {Phys. Rev. Lett.}\
  }\textbf {\bibinfo {volume} {60}},\ \bibinfo {pages} {944} (\bibinfo {year}
  {1988})}\BibitemShut {NoStop}%
\bibitem [{\citenamefont {Schrieffer}\ \emph {et~al.}(1989)\citenamefont
  {Schrieffer}, \citenamefont {Wen},\ and\ \citenamefont
  {Zhang}}]{Schrieffer1989}%
  \BibitemOpen
  \bibfield  {author} {\bibinfo {author} {\bibfnamefont {J.~R.}\ \bibnamefont
  {Schrieffer}}, \bibinfo {author} {\bibfnamefont {X.-G.}\ \bibnamefont {Wen}},
  \ and\ \bibinfo {author} {\bibfnamefont {Z.-C.}\ \bibnamefont {Zhang}},\
  }\href@noop {} {\bibfield  {journal} {\bibinfo  {journal} {Phys. Rev. B}\
  }\textbf {\bibinfo {volume} {39}},\ \bibinfo {pages} {11663} (\bibinfo {year}
  {1989})}\BibitemShut {NoStop}%
\bibitem [{\citenamefont {Jagla}\ \emph {et~al.}(1993)\citenamefont {Jagla},
  \citenamefont {Hallberg},\ and\ \citenamefont {Balseiro}}]{Jagla1993}%
  \BibitemOpen
  \bibfield  {author} {\bibinfo {author} {\bibfnamefont {E.~A.}\ \bibnamefont
  {Jagla}}, \bibinfo {author} {\bibfnamefont {K.}~\bibnamefont {Hallberg}}, \
  and\ \bibinfo {author} {\bibfnamefont {C.~A.}\ \bibnamefont {Balseiro}},\
  }\href {\doibase 10.1103/PhysRevB.47.5849} {\bibfield  {journal} {\bibinfo
  {journal} {Phys. Rev. B}\ }\textbf {\bibinfo {volume} {47}},\ \bibinfo
  {pages} {5849} (\bibinfo {year} {1993})}\BibitemShut {NoStop}%
\bibitem [{\citenamefont {Kollath}\ \emph {et~al.}(2005)\citenamefont
  {Kollath}, \citenamefont {Schollw\"ock},\ and\ \citenamefont
  {Zwerger}}]{Kollath2005}%
  \BibitemOpen
  \bibfield  {author} {\bibinfo {author} {\bibfnamefont {C.}~\bibnamefont
  {Kollath}}, \bibinfo {author} {\bibfnamefont {U.}~\bibnamefont
  {Schollw\"ock}}, \ and\ \bibinfo {author} {\bibfnamefont {W.}~\bibnamefont
  {Zwerger}},\ }\href {\doibase 10.1103/PhysRevLett.95.176401} {\bibfield
  {journal} {\bibinfo  {journal} {Phys. Rev. Lett.}\ }\textbf {\bibinfo
  {volume} {95}},\ \bibinfo {pages} {176401} (\bibinfo {year}
  {2005})}\BibitemShut {NoStop}%
\bibitem [{\citenamefont {Cheneau}\ \emph {et~al.}(2012)\citenamefont
  {Cheneau}, \citenamefont {Barmettler}, \citenamefont {Poletti}, \citenamefont
  {Endres}, \citenamefont {Schau{\ss}}, \citenamefont {Fukuhara}, \citenamefont
  {Gross}, \citenamefont {Bloch}, \citenamefont {Kollath},\ and\ \citenamefont
  {Kuhr}}]{Cheneau2012}%
  \BibitemOpen
  \bibfield  {author} {\bibinfo {author} {\bibfnamefont {M.}~\bibnamefont
  {Cheneau}}, \bibinfo {author} {\bibfnamefont {P.}~\bibnamefont {Barmettler}},
  \bibinfo {author} {\bibfnamefont {D.}~\bibnamefont {Poletti}}, \bibinfo
  {author} {\bibfnamefont {M.}~\bibnamefont {Endres}}, \bibinfo {author}
  {\bibfnamefont {P.}~\bibnamefont {Schau{\ss}}}, \bibinfo {author}
  {\bibfnamefont {T.}~\bibnamefont {Fukuhara}}, \bibinfo {author}
  {\bibfnamefont {C.}~\bibnamefont {Gross}}, \bibinfo {author} {\bibfnamefont
  {I.}~\bibnamefont {Bloch}}, \bibinfo {author} {\bibfnamefont
  {C.}~\bibnamefont {Kollath}}, \ and\ \bibinfo {author} {\bibfnamefont
  {S.}~\bibnamefont {Kuhr}},\ }\href {http://dx.doi.org/10.1038/nature10748}
  {\bibfield  {journal} {\bibinfo  {journal} {Nature}\ }\textbf {\bibinfo
  {volume} {481}},\ \bibinfo {pages} {484 EP } (\bibinfo {year}
  {2012})}\BibitemShut {NoStop}%
\bibitem [{\citenamefont {Herbrych}\ \emph {et~al.}(2017)\citenamefont
  {Herbrych}, \citenamefont {Feiguin}, \citenamefont {Dagotto},\ and\
  \citenamefont {Heidrich-Meisner}}]{Herbrych2017}%
  \BibitemOpen
  \bibfield  {author} {\bibinfo {author} {\bibfnamefont {J.}~\bibnamefont
  {Herbrych}}, \bibinfo {author} {\bibfnamefont {A.~E.}\ \bibnamefont
  {Feiguin}}, \bibinfo {author} {\bibfnamefont {E.}~\bibnamefont {Dagotto}}, \
  and\ \bibinfo {author} {\bibfnamefont {F.}~\bibnamefont {Heidrich-Meisner}},\
  }\href {\doibase 10.1103/PhysRevA.96.033617} {\bibfield  {journal} {\bibinfo
  {journal} {Phys. Rev. A}\ }\textbf {\bibinfo {volume} {96}},\ \bibinfo
  {pages} {033617} (\bibinfo {year} {2017})}\BibitemShut {NoStop}%
\bibitem [{\citenamefont {Nagaoka}(1966)}]{Nagaoka1966}%
  \BibitemOpen
  \bibfield  {author} {\bibinfo {author} {\bibfnamefont {Y.}~\bibnamefont
  {Nagaoka}},\ }\href {\doibase 10.1103/PhysRev.147.392} {\bibfield  {journal}
  {\bibinfo  {journal} {Phys. Rev.}\ }\textbf {\bibinfo {volume} {147}},\
  \bibinfo {pages} {392} (\bibinfo {year} {1966})}\BibitemShut {NoStop}%
\bibitem [{\citenamefont {Kohno}(1997)}]{Kohno1997}%
  \BibitemOpen
  \bibfield  {author} {\bibinfo {author} {\bibfnamefont {M.}~\bibnamefont
  {Kohno}},\ }\href {\doibase 10.1103/PhysRevB.56.15015} {\bibfield  {journal}
  {\bibinfo  {journal} {Phys. Rev. B}\ }\textbf {\bibinfo {volume} {56}},\
  \bibinfo {pages} {15015} (\bibinfo {year} {1997})}\BibitemShut {NoStop}%
\bibitem [{\citenamefont {Martins}\ \emph {et~al.}(1999)\citenamefont
  {Martins}, \citenamefont {Gazza},\ and\ \citenamefont
  {Dagotto}}]{Martins1999}%
  \BibitemOpen
  \bibfield  {author} {\bibinfo {author} {\bibfnamefont {G.~B.}\ \bibnamefont
  {Martins}}, \bibinfo {author} {\bibfnamefont {C.}~\bibnamefont {Gazza}}, \
  and\ \bibinfo {author} {\bibfnamefont {E.}~\bibnamefont {Dagotto}},\
  }\href@noop {} {\bibfield  {journal} {\bibinfo  {journal} {Phys. Rev. B}\
  }\textbf {\bibinfo {volume} {59}},\ \bibinfo {pages} {13596 } (\bibinfo
  {year} {1999})}\BibitemShut {NoStop}%
\bibitem [{\citenamefont {Martins}\ \emph
  {et~al.}(2000{\natexlab{b}})\citenamefont {Martins}, \citenamefont {Gazza},
  \citenamefont {Xavier}, \citenamefont {Feiguin},\ and\ \citenamefont
  {Dagotto}}]{Martins2000a}%
  \BibitemOpen
  \bibfield  {author} {\bibinfo {author} {\bibfnamefont {G.~B.}\ \bibnamefont
  {Martins}}, \bibinfo {author} {\bibfnamefont {C.}~\bibnamefont {Gazza}},
  \bibinfo {author} {\bibfnamefont {J.~C.}\ \bibnamefont {Xavier}}, \bibinfo
  {author} {\bibfnamefont {A.}~\bibnamefont {Feiguin}}, \ and\ \bibinfo
  {author} {\bibfnamefont {E.}~\bibnamefont {Dagotto}},\ }\href@noop {}
  {\bibfield  {journal} {\bibinfo  {journal} {Phys. Rev. Lett.}\ }\textbf
  {\bibinfo {volume} {84}},\ \bibinfo {pages} {5844 } (\bibinfo {year}
  {2000}{\natexlab{b}})}\BibitemShut {NoStop}%
\bibitem [{\citenamefont {Martins}\ \emph
  {et~al.}(2000{\natexlab{c}})\citenamefont {Martins}, \citenamefont {Xavier},
  \citenamefont {Gazza}, \citenamefont {Vojta},\ and\ \citenamefont
  {Dagotto}}]{Martins2000b}%
  \BibitemOpen
  \bibfield  {author} {\bibinfo {author} {\bibfnamefont {G.~B.}\ \bibnamefont
  {Martins}}, \bibinfo {author} {\bibfnamefont {J.~C.}\ \bibnamefont {Xavier}},
  \bibinfo {author} {\bibfnamefont {C.}~\bibnamefont {Gazza}}, \bibinfo
  {author} {\bibfnamefont {M.}~\bibnamefont {Vojta}}, \ and\ \bibinfo {author}
  {\bibfnamefont {E.}~\bibnamefont {Dagotto}},\ }\href {\doibase
  10.1103/PhysRevB.63.014414} {\bibfield  {journal} {\bibinfo  {journal} {Phys.
  Rev. B}\ }\textbf {\bibinfo {volume} {63}},\ \bibinfo {pages} {014414}
  (\bibinfo {year} {2000}{\natexlab{c}})}\BibitemShut {NoStop}%
\bibitem [{\citenamefont {{Nozi\`eres, P.}}(1985)}]{Nozieres1985}%
  \BibitemOpen
  \bibfield  {author} {\bibinfo {author} {\bibnamefont {{Nozi\`eres, P.}}},\
  }\href {\doibase 10.1051/anphys:0198500100101900} {\bibfield  {journal}
  {\bibinfo  {journal} {Ann. Phys. Fr.}\ }\textbf {\bibinfo {volume} {10}},\
  \bibinfo {pages} {19} (\bibinfo {year} {1985})}\BibitemShut {NoStop}%
\bibitem [{\citenamefont {Raczkowski}\ and\ \citenamefont
  {Assaad}(2013)}]{Raczkowski2013}%
  \BibitemOpen
  \bibfield  {author} {\bibinfo {author} {\bibfnamefont {M.}~\bibnamefont
  {Raczkowski}}\ and\ \bibinfo {author} {\bibfnamefont {F.~F.}\ \bibnamefont
  {Assaad}},\ }\href@noop {} {\bibfield  {journal} {\bibinfo  {journal} {Phys.
  Rev. B}\ }\textbf {\bibinfo {volume} {88}},\ \bibinfo {pages} {085120}
  (\bibinfo {year} {2013})}\BibitemShut {NoStop}%
\end{thebibliography}

%

\end{document}